\documentclass[preprint,aps,floatfix,showpacs]{revtex4}       
\usepackage{graphicx}
\begin{document}

%
\title{Influence of multiphonon excitations and transfer\\
on the fusion of Ca+Zr}
\author{H. Esbensen$^1$ and A. M. Stefanini$^2$}
\affiliation{$^1$Physics Division, Argonne National Laboratory, Argonne, Illinois 60439}
\affiliation{
$^2$INFN, Laboratori Nazionale di Legnaro, I-35020 Legnaro (Padova), Italy}  
\date{\today}
\begin{abstract}
Fusion data for $^{48}$Ca+$^{90,96}$Zr 
are analyzed by coupled-channels calculations that are based on the
M3Y+repulsion, double-folding potential. By applying a previously 
determined  nuclear density of $^{48}$Ca, the neutron densities of 
the zirconium isotopes are adjusted to optimize the fit to the 
fusion data, whereas the proton densities are determined by electron 
scattering experiments.
It is shown that the fusion data can be explained fairly well by 
including couplings to one- and two-phonon excitations of the 
reacting nuclei and to one- and two-nucleon transfer reactions but 
there is also some sensitivity to multiphonon excitations.
The neutron skin thicknesses extracted for the two zirconium isotopes 
are consistent with anti-proton measurements.
The densities of the zirconium isotopes are used together with the 
previously determined nuclear density of $^{40}$Ca to calculate the 
M3Y+repulsion potentials and predict the fusion cross sections of 
$^{40}$Ca+$^{90,96}$Zr. 
The predicted cross sections for $^{40}$Ca+$^{90}$Zr are in reasonable 
agreement with the data when the influence of multiphonon excitations 
and a modest transfer is considered.  
The prediction of the $^{40}$Ca+$^{96}$Zr fusion cross section, on 
the other hand, is poor and under-predicts the data by 30 to 40\%.
Although couplings to transfer channels with positive $Q$ values 
were expected to play an important role, they are not able to explain 
the data, primarily because the predicted Coulomb barrier is about 
1.5 MeV too high. Possible reasons for this failure are discussed.
\end{abstract}
\pacs{21.30.Fe,24.10.Eq, 25.60.Pj, 25.70.Jj}
\maketitle

\section{Introduction}

Heavy-ion fusion reactions have been studied extensively over the past 
two decades. One of the early goals was to explain the barrier 
distributions that have been extracted from measurements \cite{dasgupta}.
Another interesting and somewhat related question is what is the influence 
of transfer. This issue was first raised by Beckerman et al. 
\cite{beckerman} who observed a large enhancement in the subbarrier fusion 
of $^{58}$Ni+$^{64}$Ni compared to an interpolation between the fusion 
cross sections for the two symmetric systems, $^{58}$Ni+$^{58}$Ni and
$^{64}$Ni+$^{64}$Ni.
A similar observation was made in the fusion of $^{40}$Ca+$^{48}$Ca 
\cite{jiangca} which is strongly enhanced at subbarrier energies and
suppressed at high energies compared to the expectation based on the 
fusion data for  the two symmetric systems, $^{48}$Ca+$^{48}$Ca 
\cite{stef4848} and $^{40}$Ca+$^{40}$Ca \cite{mon4040}.

It was realized early on that the influence of transfer on fusion can be 
large when the effective $Q$ value for transfer is large and positive 
\cite{ricardo83}, as is the case for $^{58}$Ni+$^{64}$Ni and 
$^{40}$Ca+$^{48}$Ca. 
Couplings to transfer channels with large negative $Q$  values, on the other 
hand, are expected to have a much smaller effect. 
The coupling to such reaction channels results in an energy shift of the 
calculated cross section, due to an adiabatic renormalization of the ion-ion 
potential \cite{taki12}. However, it can be difficult to disentangle this 
effect from uncertainties in the ion-ion potential and the effect is 
therefore often ignored.

A more recent observation is that fusion data are often hindered at extreme 
sub barrier energies \cite{NiY} when compared to coupled-channels calculations 
that use the standard or empirical Woods-Saxon potential which has been 
extracted from elastic scattering data \cite{AW95}. 
Another observation is that the data are often suppressed at energies far 
above the Coulomb barrier when compared to calculations that use the same 
potential \cite{newton}.  Both phenomena and ways to explain them were 
discussed in a recent review of heavy-ion fusion reactions \cite{RMP}.

Excellent examples on all of the above phenomena are observed in the 
data for the fusion of $^{40}$Ca with $^{90,96}$Zr \cite{timca40zr} 
and $^{48}$Ca with $^{90,96}$Zr \cite{stefca48zr}.  
The fusion data for $^{40}$Ca+$^{96}$Zr \cite{timca40zr} were recently  
supplemented with new measurements \cite{stef4096} that went down to a
lowest cross section of 2.4 $\pm$ 1.7 $\mu$b. The data for this system are 
particularly interesting because they are strongly enhanced at subbarrier 
energies compared with the data of the other Ca+Zr systems. Moreover,
they do not show any sign of a fusion hindrance at the lowest energies.
These features are ascribed to the influence of couplings to transfer 
channels which is expected to be strong because the effective $Q$ values 
for one- and two-nucleon transfer are all positive. The influence of 
transfer on the fusion of $^{48}$Ca+$^{96}$Zr, on the other hand, is 
expected to be small because all of the effective $Q$ values for 
transfer are large and negative.

In view of the complexity of the Ca+Zr fusion data, it is of great interest 
to perform a systematic coupled-channels calculations analysis of the fusion 
data for the four systems. The basic calculations presented here include 
couplings to one- and two-phonon excitations as well as mutual excitations 
of the low-lying surface modes in the reacting nuclei. 
More complex calculations that include couplings of up to three-phonon 
excitations and to one- and two-nucleon transfer channels are also 
performed. The motivation is to explore their influence on fusion and 
try to isolate the influence of transfer from other reaction channels.

The simplest case of the four Ca+Zr systems mentioned above is possibly the 
fusion of $^{48}$Ca+$^{96}$Zr, primarily because the effective $Q$ values 
for one- and two-nucleon transfer are all negative, and the influence 
of transfer is therefore expected to be weak. The fusion data for this 
system have already been explained fairly well by coupled-channels 
calculations \cite{esb4896} that used the M3Y+repulsion (M3Y+rep) 
potential and included couplings to one- and two-phonon excitations of
the low-lying surface modes. The potential was calculated using the 
double-folding technique \cite{misiprc} which is also used in this work.

The nuclear densities of the two calcium isotopes that are used were 
determined previously in Refs. \cite{mon4040,esb4848} by analyzing 
the fusion data for $^{40}$Ca+$^{40}$Ca \cite{mon4040} and 
$^{48}$Ca+$^{48}$Ca \cite{esb4848}.
The nuclear density parameters of the two Zr isotopes are determined 
by analyzing the data for the $^{48}$Ca induced fusion reactions with 
$^{90}$Zr and $^{96}$Zr that were measured by Stefanini et al. 
\cite{stefca48zr}. 
Having determined these parameters, one can then predict the ion-ion 
potentials and the cross sections for the fusion of $^{40}$Ca
with $^{90}$Zr and $^{96}$Zr. The quality of the predictions is
tested by comparing to the data of Timmers et al. \cite{timca40zr}
and Stefanini et al. \cite{stef4096}. 

The basic ingredients of the coupled-channels technique are summarized 
in Section II. The analysis of the data for the $^{48}$Ca induced fusion 
with $^{90}$Zr and $^{96}$Zr is presented in Section III. 
The predicted cross sections for the fusion of $^{40}$Ca with the two 
zirconium isotopes are compared to the data in Section IV, and
Section V contains the conclusions.

\section{Coupled-channels calculations}

The coupled-channels calculations that are presented are similar 
to those that were performed, for example, in the analysis of the 
fusion data for the calcium isotopes \cite{mon4040,esb4848} and 
for $^{48}$Ca+$^{96}$Zr \cite{esb4896}.
The main ingredients that are relevant to the discussion here are 
summarized below. The radii of the reacting nuclei are expressed as 
$R_i = R_{i}^{(0)} + \delta R_i$, where 
\begin{equation}
\delta R_i = R_{i}^{(0)} \sum_{n\lambda\mu} \alpha_{in\lambda\mu} 
Y^*_{\lambda\mu}({\hat r}),
\label{rad}
\end{equation}
is the nuclear surface distortion, $\alpha_{in\lambda\mu}$ are the dynamic 
deformation amplitudes, and ${\hat r}$ is a unit vector along the line between
the centers of the reacting nuclei. 

In this work the iso-centrifugal approximation \cite{ronjohn} is adopted 
in order to reduce the number of coupled equations. This approximation is 
equivalent to the rotating frame approximation \cite{tami,esblan} in which 
the z-axis points along the direction of ${\hat r}$. The spherical 
harmonics that appear in Eq. (\ref{rad}) can therefore be replaced by 
$Y_{\lambda\mu}^*({\hat r})$ = $\delta_{\mu,0} \sqrt{(2\lambda+1)/(4\pi)}$.
That implies that the magnetic quantum numbers of the reacting nuclei are 
conserved.  Since the magnetic quantum numbers of even-even nuclei are zero 
in the entrance channel, they will remain zero throughout the reaction.
 
The off-diagonal matrix elements of the surface distortion, Eq. (\ref{rad}), 
that connect the ground state $|0\rangle$ to an excited state 
$|in\lambda 0\rangle$ in nucleus $i$, has the following form
in the rotating frame approximation (see Refs. \cite{esbmph1,alge})
\begin{equation}
\langle in\lambda 0| \delta R_i |0\rangle  = 
\frac{\beta_{i n\lambda} R_i^{(0)}}{\sqrt{4\pi}}.
\label{surf}
\end{equation}
Here $\lambda$ is the multipolarity of the excitation and 
$\beta_{in\lambda}$ is the deformation parameter.
There are actually two types of deformation parameters, one associated
with the nuclear induced excitation, $\beta_{n\lambda}^N$,
and one with Coulomb excitation, $\beta_{n\lambda}^C$.

The nuclear induced excitations are generated in this work by an expansion
of the nuclear field up to second order in the total nuclear surface 
distortion, $\delta R$ =  $\delta R_1$ + $\delta R_2$.  The nuclear 
interaction is therefore approximated by \cite{esbmph1,alge,esbmph2} 
\begin{equation}
V_N = U(r) - \frac{dU(r)}{dr} \delta R 
 + \frac{1}{2} \frac{d^2U(r)}{dr^2} 
 \Bigl( (\delta R)^2 - \langle 0| (\delta R)^2 | 0\rangle\rangle\Bigr),
\label{2ndorder}
\end{equation}
where $U(r)$ is the ion-ion potential, which is the nuclear interaction in 
the entrance channel. The expression (\ref{2ndorder}) has been constructed 
in such a way that the ground state expectation value 
$\langle 0| V_N|0\rangle$ is identical to the ion-ion potential $U(r)$. 
The off-diagonal matrix elements of the nuclear interaction 
can be generated in the harmonic approximation from matrix elements of 
the form given in Eq. (\ref{surf}). 

The Coulomb excitation is usually described by an expansion of the Coulomb 
field up to  first order in the surface deformation, because it has been 
shown by Hagino et al.  \cite{hagino97} that higher-order corrections to 
the Coulomb field can safely be ignored.
The expression for the Coulomb field is therefore \cite{esbmph1,alge},
\begin{equation}
V_C = \frac{Z_1Z_2e^2}{r} +
\frac{Z_1Z_2e^2}{r}
\sum_{i n \lambda}
\frac{3}{2\lambda+1} \ 
\Bigl(\frac{R_i^C}{r}\Bigr)^\lambda \
\sum_{\mu} \alpha_{in\lambda\mu} Y^*_{\lambda\mu}({\hat r}),
\label{coul}
\end{equation}
where $R_{i}^C$ = 1.20 $A_i^{1/3}$ is the Coulomb radius of nucleus $i$.
Matrix elements of the Coulomb interaction (\ref{coul}) are 
determined by matrix elements of the surface deformation \cite{alge},
\begin{equation}
\langle in\lambda | \alpha_{in\lambda\mu}| 0 \rangle
= \frac{\beta_{in\lambda}^C}{\sqrt{2\lambda+1}},
\end{equation}
which are here expressed in terms of the Coulomb deformation parameters 
$\beta_{n\lambda}^{\rm C}$.  These parameters can be obtained from the 
so-called reduced transition probability $B(E\lambda)$ that can be
found, for example, in Refs. \cite{NDS,spear02}.
The relation between the two quantities is \cite{alge}
\begin{equation}
B(E\lambda, 0\rightarrow n\lambda) = 
(\frac{3Ze\beta_{n \lambda}^C \ R^\lambda}{4\pi})^2.
\end{equation}

The nuclear deformation parameter $\beta_{n\lambda}^{\rm N}$ is often 
assumed to be identical to the Coulomb deformation parameter simply 
because other information is not available. However, the nuclear deformation
parameters have in some cases been determined by analyzing the angular 
distributions for inelastic scattering.

Fusion data of lighter and medium heavy systems can often be explained
fairly well by coupled-channels calculations that are based on the 
second-order nuclear (\ref{2ndorder}) and the first-order Coulomb 
(\ref{coul}) interactions and include up to two-phonon excitations. 
This model is also the starting point of the calculations performed in 
this work. However, it is necessary to consider higher-order couplings 
and include higher multiphonon excitations if one wants to explain the 
fusion data of heavy and soft systems (see Refs. \cite{esbmph2,hagino97}.) 
The influence of up to three-phonon excitations will therefore be 
explored but the expansion of the nuclear interaction, 
Eq. (\ref{2ndorder}), will still be truncated at the second-order level. 
 
The fusion cross section is primarily determined by the ingoing flux obtained
from the ingoing-wave boundary conditions that are imposed at the minimum of
the pocket in the entrance channel potential. This definition is supplemented 
with the absorption in a weak and short-ranged imaginary potential,
\begin{equation}
W(r) = \frac{- W_0}{1+\exp((r-R_w)/a_w)}, 
\label{imag}
\end{equation}
where $R_w$ is the position of the pocket in the entrance channel potential.
The diffuseness $a_w$ is set to 0.2 fm, whereas the strength $W_0$ is set to 
either 2 MeV or 5 MeV as explained in the description of the calculations.
The strength of the short-ranged imaginary potential is not a serious issue 
when a standard Woods-Saxon potential is used because the calculated cross 
sections are relatively insensitive to the value of $W_0$. 
It is a more delicate issue when the shallow M3Y+rep potential is used.

\subsection{The ion-ion potential}

The coupled-channels calculations are based on the M3Y+repulsion (M3Y+rep) 
potential, the calculation of which is described in Ref.  \cite{misiprc}. 
The M3Y potential alone, $U_{\rm M3Y}$, is calculated using 
the double-folding expression,
\begin{equation} 
U_{\rm M3Y}({\bf r}) = \int d{\bf r}_1\int d{\bf r}_2 \ \rho_1({\bf r}_1) \ 
\rho_2({\bf r}_2) \ v_{\rm M3Y}({\bf r}+{\bf r}_2-{\bf r}_1),
\label{m3y}
\end{equation}
where $\rho_i(r)$ are the nuclear densities of the reacting nuclei, 
and $v_{\rm M3Y}(r)$ is the effective M3Y (direct + exchange) interaction.
The densities are parametrized in terms of the symmetrized fermi function
introduced in Ref. \cite{esbopb} with a fixed diffuseness and an adjustable
radius. 

The M3Y potential $U_{M3Y}$ is extremely deep for overlapping nuclei and
it produces a pocket in the entrance channel potential that is deeper than
the ground state energy of the compound nucleus. This unphysical condition
is repaired by introducing a repulsive potential.
The repulsive part $U_{\rm rep}$ of the M3Y+rep potential \cite{misiprc},
$U_{\rm M3Y}+U_{\rm rep}$, is calculated  from an expression that is similar 
to Eq. (\ref{m3y}),
\begin{equation} 
U_{\rm rep}({\bf r}) = \int d{\bf r}_1\int d{\bf r}_2 \ 
{\hat \rho}_1({\bf r}_1) \ {\hat \rho}_2({\bf r}_2) \ 
v_{\rm rep}({\bf r}+{\bf r}_2-{\bf r}_1).
\label{rep}
\end{equation}
Here the effective interaction $v_{\rm rep}$ is assumed to be a simple
contact interaction,
\begin{equation}
v_{\rm rep}({\bf r}) = V_{\rm rep} \delta({\bf r}).
\end{equation}
The nuclear densities ${\hat \rho}_i(r)$ that are used to calculate the 
repulsive potential, Eq. (\ref{rep}), are assumed to have the same radii 
as those that are used to calculate the M3Y potential, $U_{\rm M3Y}$, but 
the diffuseness $a_{r}$ is different and is treated as an adjustable 
parameter \cite{misiprc}.

The strength of the repulsive interaction $V_{\rm rep}$ is calibrated
(once the radii and value of $a_{r}$ has been chosen) to produce a nuclear 
incompressibility $K$ of the compound nucleus that is consistent with 
the values tabulated in the work of Myers and \' Swi\c atecki \cite{myers}. 
The procedure is explained in detail in Ref. \cite{misiprc}. 
There are therefore two adjustable parameters for each of the reacting 
nuclei, namely, the radius of the density and the diffuseness associated 
with the repulsive part of the interaction. 
While the height of the Coulomb barrier is primarily determined by the
radius parameter, the diffuseness parameter $a_r$ controls the thickness 
of the barrier and the depth of the pocket in the entrance channel potential. 
Both parameters are adjusted to optimize the fit to the fusion data. 
This scheme was used in the analysis of the fusion data for the two 
symmetric systems, $^{40}$Ca+$^{40}$Ca \cite{mon4040} and 
$^{48}$Ca+$^{48}$Ca \cite{esb4848}, and the parameters of the 
densities that were obtained are shown in Table II.

By adopting the density of $^{48}$Ca that was obtained in a previous work
\cite{esb4848} one can now analyze the $^{48}$Ca+$^{90,96}$Zr fusion data 
and determine the density parameters of the two zirconium nuclei.
Actually, since the point-proton densities are fairly well known from 
electron scattering experiments, it is better to adopt these densities and 
instead calibrate the point-neutron densities to provide an optimum fit to the 
fusion data. The parameters that have been obtained are shown in Table II.
The results suggest that both isotopes have a neutron skin, which is 
defined as the difference between the RMS radii if the point-neutron 
and point-proton densities,
\begin{equation}
\delta r_{np} = \langle r^2\rangle^{1/2}_n - \langle r^2\rangle^{1/2}_p.
\label{skin}
\end{equation}

As a consistency check of the analysis, one can compare the neutron skin
thickness extracted from the fusion data to the values that have been
obtained in other experiments. Thus, if the extracted neutron skin is too
thick, that would indicate that some of the couplings were too weak or
that some important reaction channels were missing in the coupled-channels
calculations. On the other hand, if the extracted neutron skin is too thin,
that would indicate that the coupling strengths to certain reaction channels
were too strong.

We have chosen in this work to compare the extracted neutron skin
thikness to the values that have been obtained in anti-proton 
experiments \cite{trzcinska}. One reason is that systematic 
results have been obtained with this method for a wide range of nuclei. 
Another reason is that the uncertainties of this method are not 
unreasonable in comparison to other methods.
A better value could possibly be obtained by measuring the 
parity-violating asymmetry in the elastic scattering of polarized 
electrons but that has not yet been achieved \cite{pvasym}.

The values of the neutron skin thickness obtained with different 
methods are compared in Fig. 7 of Ref. \cite{dipole} for a $^{208}$Pb
target. The figure shows that the anti-proton experiment \cite{trzcinska}
($\delta_{np}$ = 0.15 $\pm$ 0.02 fm) is consistent with 
a measurement of the dipole polarizability obtained in inelastic proton 
scattering \cite{dipole} ($\delta r_{np}$ = 0.165 $\pm$ 0.026 fm). 
The elastic proton scattering data give slightly larger values of the 
neutron skin thickness but a previous analysis at 800 MeV \cite{LRay} 
gave a result ($\delta r_{np}$ = 0.14 $\pm$ 0.04 fm) that is close to 
the value of the anti-proton experiment. 
Thus it appears that proton and anti-proton experiments are in 
fairly good agreement for $^{208}$Pb, and a similar result 
($\delta_{np}$ = 0.16 $\pm$ 0.04 fm) has also been obtained in 
($^3$He,$t$) charge-exchange reactions \cite{zegers}.
The parity-violating asymmetry experiment \cite{pvasym}, on the 
other hand, gave a larger 
and more uncertain value, $\delta r_{np}$ = $0.33_{-0.18}^{+0.16}$ fm.

\subsection{Nuclear structure input}


The deformation parameter for Coulomb and nuclear induced excitations,
$\beta_\lambda^C$ and $\beta_\lambda^N$, respectively, are not always 
identical as discussed in Sec. II. An example is the excitation of the 
calcium isotopes \cite{flem}, where an analysis of the 
$^{16}$O+$^{40,48}$Ca elastic and inelastic scattering data gave 
nuclear deformation parameters that were significantly smaller than 
the adopted values for Coulomb excitation.
The two types of deformation parameters are compared in Table I in 
terms of the quantity
\begin{equation}
\sigma_\lambda = \frac{\beta_\lambda R}{\sqrt{4\pi}},
\end{equation}
which is just the matrix elements of the surface distortion, Eq. (2).
The nuclear deformation parameters for the two zirconium isotopes are 
not known so they are assumed to be identical to the Coulomb deformation
parameters.

The basic structure input to the calculations is summarized in Table I. 
When excitations of multiphonon states are considered, it is assumed that 
the couplings to these states can be calculated in the harmonic oscillator 
model from the values of $\beta_\lambda$ that describe the one-phonon 
excitation.  It must be emphasized that this approximation may not always 
be realistic and the calculations that include multiphonon excitations 
may therefore be uncertain. 

The basic two-phonon calculation includes  
one- and two-phonon excitations of the low-lying $2^+$ and $3^-$ states 
in projectile and target, as well as mutual excitations of these
states. That results in a total of 15 channels (including 
the elastic) and is referred to as the Ch-15 calculation.
The $5^-$ state in $^{48}$Ca is relatively weak and is ignored. 
The $5^-$ excitation in $^{90}$Zr is stronger but it is 
combined with the 3$^-$ excitation in the same nucleus into one
effective $3^-$ channel. The basic two-phonon calculation performed 
for the $^{48}$Ca induced fusion with $^{90}$Zr and ${96}$Zr are 
therefore Ch-15 calculations and they are reported in Section III. 

The $5^-$ excitation in $^{40}$Ca is relatively strong and it is 
therefore included explicitly in the calculations of the 
$^{40}$Ca induced fusion reactions with the zirconium isotopes.
On the other hand, the two-phonon excitations of the $2^+$ states 
in projectile and target are relatively weak and they are therefore 
ignored, and so is the two-phonon excitation of the $5^-$ state.
The basic two-phonon calculation (that includes mutual excitations)
has 18 coupled channels, and the results of this Ch-18 calculation
are reported in section IV.

The sensitivity to multiphonon excitations is investigated  
by considering the effect of up to three-phonon excitations. 
These calculations are not meant to be complete because the 
number of channels can easily become very large. The detailed
calculations are described in Sections III and IV. 

\subsection{Transfer reactions}

The effective $Q$ value for transfer reactions is defined in terms of the 
true $Q$ value \cite{ricardo83},
\begin{equation}
Q_{eff} = Q + \Delta V_{CB},
\end{equation}
where $\Delta V_{CB}$ is difference in the height of the Coulomb barrier
in the entrance and in the exit channel. The general experience is that
couplings to transfer reactions with positive effective $Q$ values
can have a strong effect and enhance the subbarrier fusion cross section
\cite{ricardo83}, whereas the effect of transfer channels 
with large negative $Q$ values is weaker and leads 
to an adiabatic renormalization of the ion-ion potential \cite{taki12}. 

The effective $Q$ values for the most favorable one- and two-nucleon 
transfer reactions in Ca+Zr collisions are shown in Table III.  
They are all negative for $^{48}$Ca+$^{96}$Zr and couplings to transfer 
channels were ignored in the previous work \cite{esb4896}  
but they will be considered in this work.
The effective $Q$ values for some of the other isotope combinations are 
positive and their influence on fusion will therefore be considered. 
The effective $Q$ values for pair-transfer are particularly large and 
positive in $^{40}$Ca+$^{96}$Zr collisions and the influence of 
pair-transfer is expected to play an important role 
in the fusion of this system.

The influence of transfer reactions is described by the model 
introduced in Ref. \cite{esbnisn}. The model assumes that 
excitations and transfer are independent degrees of freedom. 
The two-nucleon transfer  is treated both as a direct and a 
successive process. The direct two-nucleon (or pair) transfer 
is described by the monopole form factor \cite{dasso85},
\begin{equation}
F_{2p} = - \sigma_{2p} \frac{dU}{dr}.
\label{2ptr}
\end{equation}

The parameters of the one-nucleon transfer channels are specified
later on for each of the four systems. 
It is unfortunately not feasible to consider all transfer channels.
The choice of the transfer mode considered, either proton or neutron, 
is therefore made according to which transfer mode has the larger $Q$ 
value for pair-transfer. The strength of the pair-transfer, $\sigma_{2p}$, 
is adjusted in each case to optimize the fit of the calculated 
cross section to the fusion data.  This empirical approach is adopted
here because the pair transfer strength is not always known. 
The approach could be misleading because the adjusted strength could 
very well simulate the effect of couplings to other reaction channels
that are not included explicitly in the calculation.
On the other hand, if the cross sections for one- and two-nucleon 
transfer were known experimentally, one could calibrate the strengths 
of the transfer couplings so that the transfer data were reproduced 
as it was done in Ref. \cite{esbnisn}.

The influence on fusion of a particular reaction channel is primarily 
determined by the $Q$ value of the reaction and by the strength of 
the coupling at the Coulomb barrier according to the constant 
coupling model of Dasso et al. \cite{dasso83}.
The influence of the direct two-nucleon transfer is therefore 
expected to be stronger than the influence of one-nucleon transfer 
because the pair-transfer form factor, Eq. (\ref{2ptr}), is short 
ranged and relatively large at the Coulomb barrier, whereas the 
one-nucleon transfer form factor has a longer range and is relatively 
weak at the location of the Coulomb barrier. 
The cross section for one-nucleon transfer may very well be larger
than the cross section for the two-nucleon transfer but that does
not necessarily imply that the one-nucleon transfer has a large
impact on fusion.

The couplings to one- and two-nucleon transfer are included in the 
coupled-channels calculations as described in the model of 
Ref. \cite{esbnisn}. The model assumes as mentioned earlier that 
excitations and transfer are independent 
degrees of freedom. Thus, if there are 15 elastic and inelastic 
channels in the entrance channel mass partition, then the number is 
30 channels (Ch-30) when one-nucleon transfer is included, and 45 
channels (Ch-45) when both one-and two-nucleon transfer is considered.

\section{Analysis of $^{48}$Ca induced fusion}

The $^{48}$Ca+$^{90,96}$Zr fusion data \cite{stefca48zr} are analyzed
by coupled-channels calculations that use the M3Y+rep, double-folding 
potential. The nuclear density of $^{48}$Ca was determined previously 
\cite{esb4848} by analyzing the $^{48}$Ca+$^{48}$Ca fusion data 
\cite{stef4848} and it is used here to determine the densities of the 
two zirconium isotopes that provide the best fit to the 
$^{48}$Ca+$^{90,96}$Zr fusion data.
The repulsive part of the M3Y+rep interaction is calibrated to produce 
the incompressibilities $K$ = 227.9 and 223.7 MeV, respectively, that 
are predicted for the two compound nuclei, $^{138}$Nd and $^{144}$Nd 
\cite{myers}.
The overall systematic uncertainty of the experiment \cite{stefca48zr} is 
14\%, whereas the relative errors are mainly determined by statistics.
A systematic error of 5\% is therefore adopted in the $\chi^2$ analysis 
of the data. 

The data are also analyzed by adopting the standard Woods-Saxon 
parametrization of the ion-ion potential \cite{AW95}. The best fit 
to the data is achieved by adjusting the radius of the Woods-Saxon well.
Some of the results are reported here because they provide useful 
information when compared to the results obtained with the M3Y+rep 
potential. Thus a large value of the $\chi^2/N$ obtained with the 
Woods-Saxon potential may indicate a hindrance of the fusion data 
at very low energies, as discussed by Jiang et al. \cite{NiY}, or 
a suppression of the fusion data at high energies, as discussed by 
Newton et al. \cite{newton}.
The $\chi^2/N$ is expected to be smaller when the M3Y+rep potential
is used because this potential usually resolves the discrepancies 
with the data both at very low and very high energies \cite{misiprc}.
However, the situation is not always so straightforward because there 
are other issues that can play a role, for example, the influence of 
couplings to multiphonon excitations and transfer reactions.

The results of the analysis of the fusion data that is based on the
M3Y+rep potential and includes couplings to one- and two-phonon 
excitations are shown in Fig. \ref{ca48zrffch15}.  
The calculations have 15 channels and are denoted Ch-15 in the figure.
The neutron densities of the two zirconium isotopes were adjusted as 
described previously to optimize the fit to the data. 
The two parameters of the neutron density that were obtained, namely, the 
radius $R_n$ and the diffuseness $a_r$ associated with the repulsion, 
are shown in Table IV and V. 
Also shown is the height $V_{CB}$ of the Coulomb
barrier in the entrance channel, as well as the quality of fit to the 
data in terms of the $\chi^2/N$.
The fits to the data in Ch-15 calculations are not perfect and it is 
discussed below how they can be improved by considering the effect of 
transfer and multiphonon excitations. 

\subsection{$^{48}$Ca+$^{90}$Zr}

\begin{figure}
\includegraphics[width = 8cm]{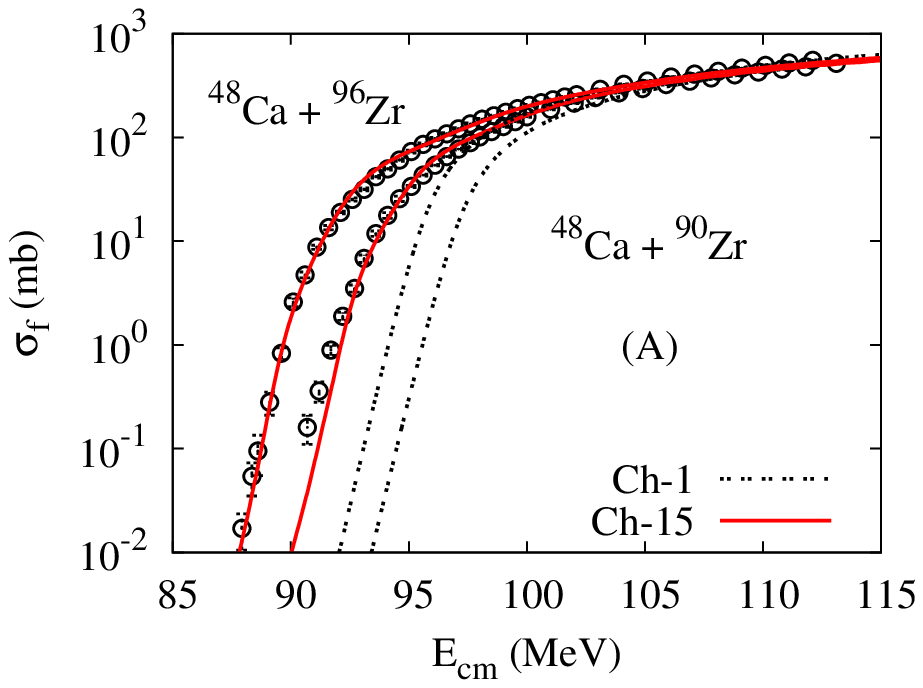}
\includegraphics[width = 8cm]{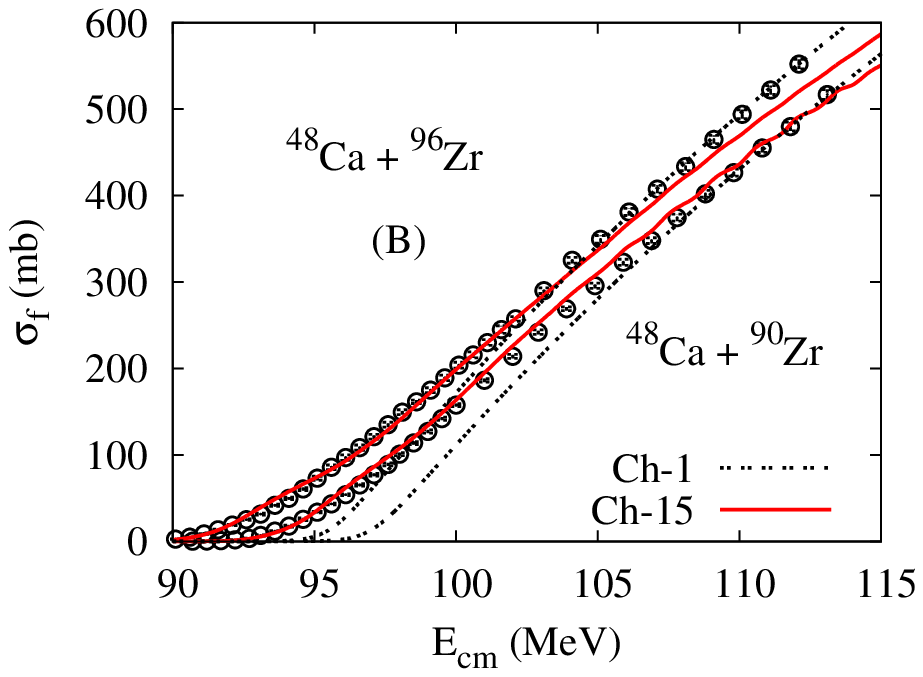}
\caption{
(Color online)
The measured cross sections for the fusion of $^{48}$Ca+$^{90,96}$Zr 
\cite{stefca48zr} are compared to Ch-15 calculations that are based on 
the M3Y+rep potential. The Ch-1 no-coupling calculations are also shown.
The cross sections are shown in a logarithmic (A) and a linear plot (B).}
\label{ca48zrffch15}
\label{ca48zrflfch15}
\end{figure}

The fit to the $^{48}$Ca+$^{90}$Zr fusion data shown in Fig. 
\ref{ca48zrffch15} is poor.
It is seen that the Ch-15 calculation underpredicts the data
at subbarrier energies (see Fig. \ref{ca48zrffch15}A) and exceeds 
them at high energies (see Fig.  \ref{ca48zrffch15}B).
It turns out that a previous analysis that used a standard 
Woods-Saxon potential with diffuseness $a$ = 0.68 fm gave 
essentially the same result \cite{stefca48zr}.  
It is shown below that it is possible to achieve a much better 
fit by including couplings to one- and two-proton transfer 
reactions. The calculation is denoted Ch-45 and contains 45 
coupled channels as explained in section II.C.

\noindent
{\it Effects of transfer.}
The best fit to the data in Ch-45 calculations is achieved by 
adjusting the strength $\sigma_{2p}$ of the two-proton transfer, Eq. 
(\ref{2ptr}), as well as the radius of the neutron density in 
$^{90}$Zr.  The effective $Q$ value for the one-proton transfer, 
from fully occupied $p_{1/2}$ orbit in $^{90}$Zr to the unoccupied 
$f_{7/2}$ proton orbit outside $^{48}$Ca, is set to -0.92 MeV,
(see Table III.) The effective $Q$ value for two-proton transfer 
is +2.22 MeV but the value $Q_{2p}$=+1 MeV is adopted here because 
it provides the best fit to the data.
The neutron transfer is ignored because the $Q$ values are 
negative so the influence on fusion is expected to be smaller.

\begin{figure}
\includegraphics[width = 8cm]{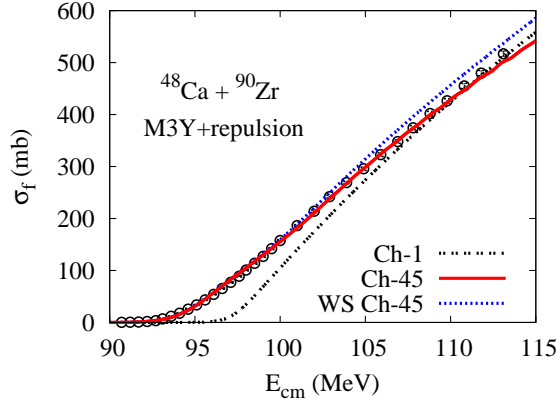}
\caption{(Color online)
Linear plot of cross sections for the fusion of  $^{48}$Ca+$^{90}$Zr 
\cite{stefca48zr}.  The best Ch-45 calculations for the Woods-Saxon (WS) and
the M3Y+rep potentials are shown. The no-coupling calculation Ch-1 based on
the M3Y+rep potential is also shown. All calculations use a short-ranged 
absorption with $W_0$=2 MeV.}
\label{4890flf}
\end{figure}

The adjusted Ch-45 calculations are in good agreement with the data, 
with a $\chi^2/N$ = 1.06 for the standard Woods-Saxon potential,
and $\chi^2/N$ = 0.69 for the M3Y+rep potential (see Table IV.) 
The two calculations are compared to the data in Fig. \ref{4890flf}. 
The calculation with the Woods-Saxon potential (WS Ch-45) exceeds 
the high energy data. This is in qualitative agreement with the 
systematics found by Newton et al. \cite{newton}.  
The calculation based on the M3Y+rep potential resolves the 
discrepancy at high energies and provides an almost perfect fit. 
The adjusted pair-transfer strength is $\sigma_{2p}$ = 0.15 fm in 
both calculations and it produces a pair-transfer cross section of 
the order of 80 mb at 110 MeV. 

\begin{figure}
\includegraphics[width = 8cm]{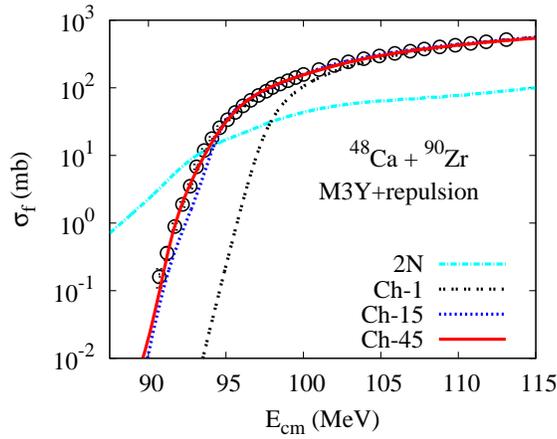}
\caption{(Color online)
Cross sections for the fusion of $^{48}$Ca+$^{90}$Zr \cite{stefca48zr}
are compared to Ch-15 and Ch-45 calculations that are based on the 
M3Y+rep potential that gives the best fit in Ch-45 calculations. 
The result of the no-coupling limit (Ch-1)
and the predicted 2N cross sections are also shown.}
\label{4890ff}
\end{figure}

The influence of transfer on the fusion of $^{48}$Ca+$^{90}$Zr is 
relatively modest at high energies, where it reduces the cross 
section slightly and brings it into better agreement with the data. 
The effect is much larger at low energies. This can be seen in Fig. 
\ref{4890ff}, where the results of Ch-15 and Ch-45 calculations are 
compared to the data. Both calculations apply the M3Y+rep potential 
that gives the best fit in Ch-45 calculations 
(see Table IV). The fit of the Ch-15 calculation is therefore not as 
good as obtained in the Ch-15 calculation shown in Fig. 1. 
It is shown here so one can see directly the effect of transfer by 
comparing the Ch-15 and Ch-45 calculations. 

\noindent
{\it Effects of multiphonon excitations.}
In order to limit the number of channels in calculations that include 
up to three-phonon excitations, we first exclude the two-phonon 
excitations of the $2^+$ states in projectile and target because the 
excitation strength is relatively weak (see Table I, where 
$\beta_2 \approx$ 0.1).  The number of channels in the basic two-phonon 
calculation is therefore reduced from 15 to 13 channels. 
This calculation is supplemented with the simultaneous excitation of 
3 different one-phonon states and with the combined excitation of a 
one-phonon and a two-phonon state. 
The explicit three-phonon excitation of any particular one-phonon 
state is ignored. The total number of channels in the $^{48}$Ca 
induced reactions is therefore 23 and the calculation is referred 
to as the Ch-23 three-phonon calculation.

The effect of couplings to three-phonon excitations turns out to be
relatively modest. 
The best Ch-23 calculation has a $\chi^2/N$ of 2.3 in contrast to the 
much smaller value of 0.69 obtained in the Ch-45 calculation discussed 
above that included the effect of transfer and two-phonon excitations.
This conclusion is confirmed by Ch-69 calculations that include the 
combined effect of transfer and up to three-phonon excitations because
the $\chi^2/N$ is the same as obtained in Ch-45 calculations (see Table IV.) 

\begin{figure}
\includegraphics[width = 8cm]{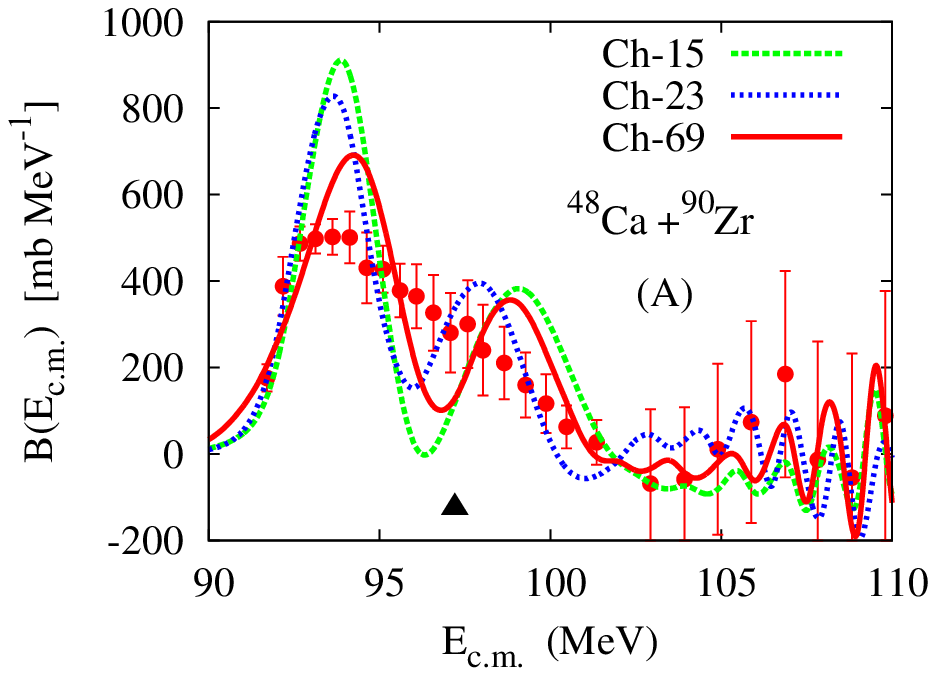}
\includegraphics[width = 8cm]{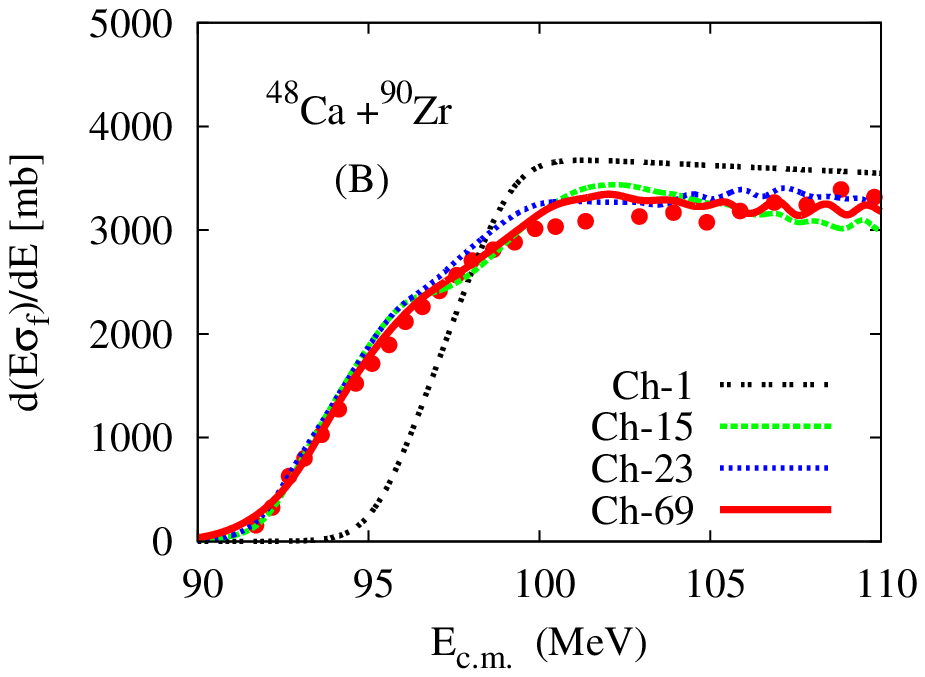}
\caption{(Color online)
Barrier distributions (A) and the first derivative (B) of the energy-weighted 
cross sections for the fusion of  $^{48}$Ca+$^{90}$Zr \cite{stefca48zr}.
The calculations used the M3Y+rep potential with the radius $R_n$ = 4.835 fm
of the neutron density in $^{90}$Zr.
The energy of the Coulomb  barrier is indicated in (A) by the large triangle.}
\label{4890f2d}
\end{figure}

Although the couplings to three-phonon excitations do not play any major 
role in the overall $\chi^2$ fit to the fusion data, their influence 
can be seen in the barrier distribution for fusion, which is defined 
as the second derivative of the energy-weighted cross section \cite{rowley},
\begin{equation}
B(E) = \frac{d^2(E\sigma)}{dE^2}.
\label{barrier}
\end{equation}
The distribution is calculated using the finite difference method with 
an energy step of $\Delta E$ = 2 MeV. 
The distributions one obtains for the fusion of $^{48}$Ca+$^{90}$Zr 
are illustrated in Fig. \ref{4890f2d}A. The Ch-15 calculation
is seen to consist of two isolated peaks, whereas the measured 
distribution is broad and asymmetric. The Ch-23 calculation shows a 
slight improvement and so does the Ch-69 calculation but the distribution 
is still dominated by two peaks.

The discrepancy with the measured barrier distribution suggests that
some important reaction mechanism is still missing, or maybe the 
reaction model used here, which assumes that excitation and transfer 
are independent degrees of freedom, is unrealistic. 
Another possibility is that couplings to a large number of 
(non-collective) excitation and transfer channels, with a wide range 
of $Q$ values, would smear the calculated barrier distribution and 
bring it into better agreement with the measurement. 
A similar hypothesis was proposed by Yusa et al.  \cite{yusa} but 
it did not explain the data in the case they studied.

The first derivative of the energy weighted cross section is shown
in Fig. \ref{4890f2d}B. By comparing the Ch-15 and Ch-23 calculations 
it is seen that the influence of multiphonon excitations is weak,
and by comparing the Ch-23 and Ch-69 it is seen that the influence 
of transfer is relatively modest at high energies.

As a final test of the convergence of the results obtained in
the analysis of the fusion data, it is useful to take a  look 
at the neutron densities that have been extracted. 
The parameters of the assumed point-proton and the extracted 
point-neutron densities of $^{90}$Zr are listed in Table II. 
From the associated RMS radii given in column 5 one can now 
determine the thickness of the neutron skin, $\delta r_{np}$,
defined in Eq. (\ref{skin}).
The skin thickness decreases as the number of channels increases,
from $\delta r_{np}$ = 0.14 fm in the Ch-15 calculation,
to $\delta r_{np}$ = 0.08 fm in the Ch-45 calculation.
The latter value is in surprisingly good agreement with the 
anti-proton experiment \cite{trzcinska} 
that gave the value $\delta r_{np}$ = 0.09 $\pm$ 0.025 fm.  

\subsection{$^{48}$Ca+$^{96}$Zr}

The Ch-15 fit to the $^{48}$Ca+$^{96}$Zr data shown in Fig.
\ref{ca48zrffch15} looks reasonable and has a $\chi^2/N$ = 1.5. 
The calculation is essentially the same that was published in Ref. 
\cite{esb4896}. There are some discrepancies at high energies 
(see the linear plot in Fig. \ref{ca48zrffch15}B), where the data 
exceed the calculation. This is the opposite of the systematics
trend discussed by Newton et al. \cite{newton}. Ways to eliminate 
the discrepancies are discussed below.

It is interesting that the $\chi^2/N$ shown in Table V are the same 
in the Ch-15 calculations that are based on the Woods-Saxon and on 
the M3Y+rep potential, respectively. This is misleading because the 
data are actually hindered at low energies compared to calculations
that use a standard Woods-Saxon potential.
This was already demonstrated in Ref.  \cite{esb4896}. One would 
therefore expect that the M3Y+rep potential would provide the best 
description to the data. The main reason this is not the case is 
that the Ch-15 calculation, which uses the M3Y+rep potential and 
a weak, short-ranged imaginary potential, underestimates the data 
at high energies.
This problem was solved in Ref. \cite{esb4896} by applying a stronger 
short-ranged imaginary potential, which should simulate the effect of
couplings to channels that were not included explicitly in the calculations. 
It would be more satisfactory if one could explain the fusion data without
resorting to a strong imaginary potential. The influence of couplings to 
multiphonon excitations and transfer reactions is therefore investigated 
below.

\noindent {\it Effects of multiphonon excitations.}
The influence of multiphonon excitations can be seen in the barrier
distribution  for fusion. This was demonstrated in the original analysis 
\cite{stefca48zr}, where it was shown that some of the structures that 
appear in the measured distribution can be explained by considering 
multiphonon excitations of the low-lying $3^-$ state in $^{96}$Zr. 
This explanation was confirmed in Ref. \cite{esb4896}.

\begin{figure}
\includegraphics[width = 8cm]{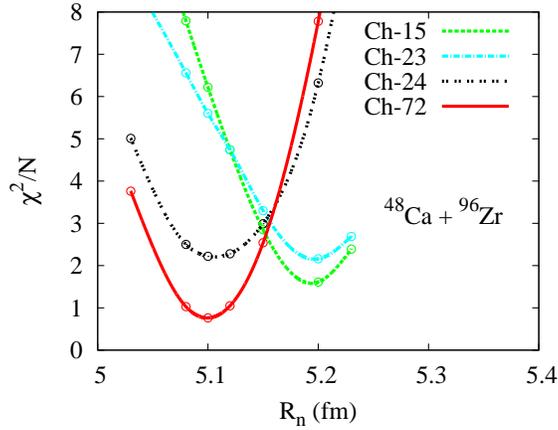}
\caption{ (color online)
The $\chi^2/N$ for the $^{48}$Ca+$^{96}$Zr fusion data \cite{stefca48zr}
obtained in Ch-15, Ch-23, Ch-24, and Ch-72 coupled-channels calculations
as function of the radius $R_n$ of the neutron density of $^{96}$Zr.}
\label{4896xki2}
\end{figure}

The cleanest way to study the influence of multiphonon excitations 
is to repeat the data analysis with an increasing number of channels 
and plot the $\chi^2/N$ as a function of the radius $R_n$ of the 
neutron density. The results of the analysis are shown in Fig. 
\ref{4896xki2}. The calculations employed the fixed value $W_0$ = 5 
MeV of the short-ranged imaginary potential, and a fixed diffuseness 
of $a_r$ = 0.395 fm  associated with the repulsive part of the M3Y+rep 
potential. The latter value gives the optimum fit to the data in the 
Ch-15 calculations discussed above.

The $\chi^2/N$ obtained in Ch-15 and Ch-23 calculations are shown in
Fig. \ref{4896xki2}. They are similar in magnitude and both have a 
minimum near $R_n$ = 5.20 fm. It is therefore concluded that the 
three-phonon excitations that are included in Ch-23 calculations
have a relatively modest influence on the fusion. 
The Ch-24 calculations which, in addition to the states of the Ch-23 
calculation, include the three-phonon excitation of the low-lying 
$3^-$ state in $^{96}$Zr, has a minimum near $R_n$ = 5.10 fm. 
That is a significant change in radius, from the 5.20 fm to 5.10 fm.
It implies that the neutron skin thickness of $^{96}$Zr is reduced 
from 0.23 fm in the Ch-15 calculation to 0.16 fm in yhe
Ch-72 calculation (see Table II.) 
The latter result is consistent with the neutron skin thickness 
$\delta r_{np}$ = 0.12 $\pm$ 0.05 fm that was obtained in the 
anti-proton experiment \cite{trzcinska}. 

\noindent {\it Influence of transfer.}
Finally, the influence of transfer channels is studied in Ch-72 
calculations that are based on the same excitation channels that 
were used in the Ch-24 calculations discussed above and include, 
in addition, the one- and two-neutron transfer. 
The neutron transfer channels are chosen here because they have 
the most favorable $Q$ values (see Table III.)
The one-neutron transfer, from the fully occupied $d_{5/2}$ orbit 
in $^{96}$Zr to the unoccupied $p_{1/2}$ state in $^{48}$Ca, has
the effective $Q$ value $Q_{1n}$ = -2.64 MeV and the two-neutron 
transfer has the $Q$ value $Q_{2n}$ = -2.67 MeV. The best fit to the
data in Ch-72 calculations is achieved for a modest pair-transfer 
strength of $\sigma_{2p}$ = 0.05 fm. 

The $\chi^2/N$ for Ch-72 calculations is shown in Fig. \ref{4896xki2}.
It has a minimum at the radius $R_n$ = 5.10 fm which gave the best 
fit in the Ch-24 calculations discussed above and is consistent with 
the measured neutron skin thickness of Ref. \cite{trzcinska}.
In other words, the influence of transfer does not affect the radius 
of the neutron density that is extracted from the analysis of the 
fusion data. However, it improves the fit to the data by reducing 
the $\chi^2/N$ considerably.

\begin{figure}
\includegraphics[width = 8cm]{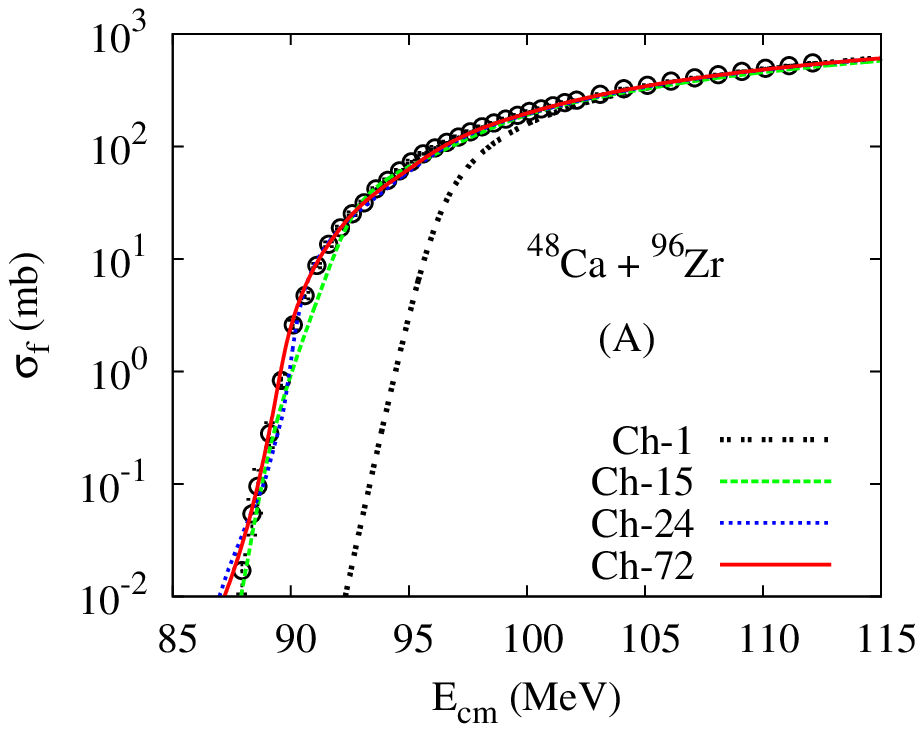}
\includegraphics[width = 8cm]{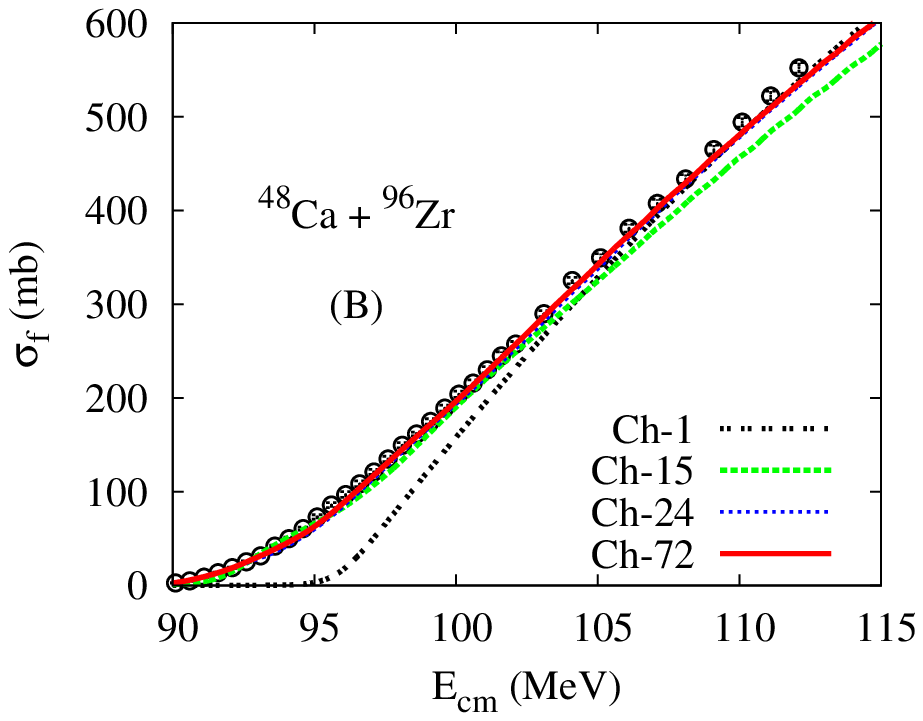}
\caption{ (color online)
Logarithmic (A) and linear plot (B) of the measured fusion cross 
sections for $^{48}$Ca+$^{96}$Zr \cite{stefca48zr} are compared to 
coupled-channels calculations that are based on the M3Y+rep potential,
obtained with a radius $R_n$ = 5.1 fm of the neutron density in $^{96}$Zr.}
\label{4896ff}
\end{figure}

The calculated cross sections, obtained with the radius $R_n$ = 5.1 fm
of the neutron density in $^{96}$Zr, are compared to the data  
in Figs. \ref{4896ff}A. It is seen that it is the combined effect of
couplings to multiphonon excitations (in Ch-24 calculations) and 
one- and two-neutron transfer reactions (in Ch-72 calculations) 
that provides the good agreement with the data at low energies.
At high energies, it is primarily the effect of the multiphonon 
excitations in Ch-24 calculations that brings the calculated cross 
section into good agreement with the data. 
This is illustrated in Fig. \ref{4896ff}B but it is difficult 
to see because the Ch-24 calculation is covered by the Ch-72 
calculation that in addition includes the effecrs of transfer.
In other words, the influence of transfer is small at high energies.

\begin{figure}
\includegraphics[width = 8cm]{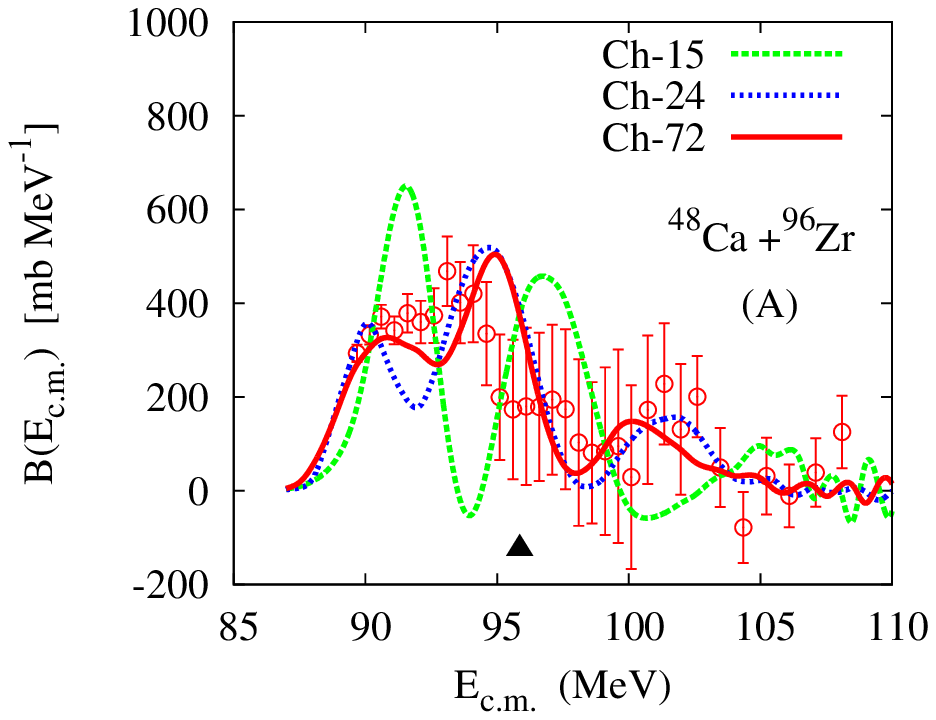}
\includegraphics[width = 8cm]{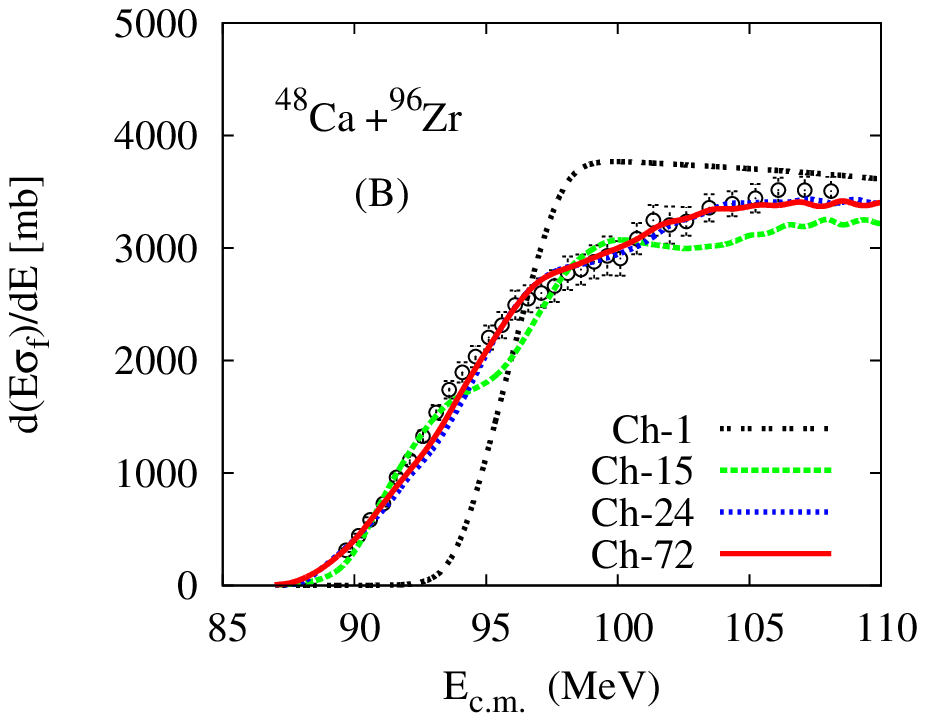}
\caption{(color online)
Barrier distributions (A) and the first derivative (B) of the 
energy-weighted fusion cross sections for $^{48}$Ca+$^{96}$Zr
shown in Fig. \ref{4896ff}.
The M3Y+rep potential used in the calculations is based on a 
$^{96}$Zr neutron density with radius $R_n$ = 5.10 fm.
The energy of the Coulomb  barrier is indicated in (A) by the large triangle.}
\label{4896f2d}
\end{figure}

\noindent {\it Derivatives of the cross section.}
At this point it is of interest to compare the barrier distribution, 
Eq. (\ref{barrier}),  extracted from the experiment to some of the 
calculations discussed above. The comparison is made in Fig. 
\ref{4896f2d}A. It is seen that the barrier distribution for the 
best Ch-15 calculation consists of two strong peaks and a broader 
bump near 105 MeV.  The Ch-24 calculation has three major peaks that 
agree fairly well with the structures that are observed in the 
measured barrier distribution.  Finally, the best Ch-72 calculation 
shows that the influence of transfer is minor but it does improve 
the shape of the distribution slightly in comparison to the measured  
barrier distribution.

A good way to illustrate the behavior of the cross sections at high 
energies is to plot the derivative of the energy-weighted cross 
sections. The measured and calculated results are shown in Fig. 
\ref{4896f2d}B. The calculations were all performed with the same 
M3Y+rep potential that provides the optimum fit to the data 
(as discussed above) in Ch-24 and Ch-72 calculations and is
determined by the neutron density with the radius $R_n$ = 5.10 fm.
It is seen that the Ch-15 calculation does a poor job in reproducing 
the data, whereas the Ch-24 and Ch-72 calculations give almost
identical results and are both in excellent agreement with the high 
energy data. 

It should be emphasized that there are uncertainties in the choice of 
the nuclear structure input to multiphonon excitations. For example, 
the multiphonon excitations are described by the harmonic oscillator 
model but that may not be a realistic assumption. It is also assumed 
that the $\beta_\lambda$ values are the same for the Coulomb and 
nuclear induced excitations of the zirconium isotopes but that is not 
necessarily a valid assumption.  On the other hand, the excellent 
agreement with the fusion data that is achieved when the three-phonon 
octupole excitation of $^{96}$Zr is included, and the consistency of 
the extracted neutron density with the measured neutron skin, is very 
encouraging. These findings will hopefully stimulate a search for 
such multiphonon excitations in nuclear structure measurements.



\section{Predictions of the $^{40}$Ca induced fusion}

The nuclear density of $^{40}$Ca that was determined in a previous 
analysis of $^{40}$Ca+$^{40}$Ca fusion data \cite{mon4040} can now 
be combined with the densities of the two zirconium isotopes to calculate 
the M3Y+rep potentials for $^{40}$Ca+$^{90,96}$Zr. 
The repulsive part of the interaction is calibrated to produce the 
incompressibilities $K$ = 232.1 and 229.1 MeV, respectively, that 
are predicted for the two compound nuclei, $^{130}$Nd and $^{136}$Nd
\cite{myers}.

The basic two-phonon calculation for the $^{40}$Ca induced fusion reactions
has 18 channels as described in Sect. II.B. 
The three-phonon calculation for the fusion of $^{40}$Ca+$^{96}$Zr that is
built on the Ch-18 two-phonon calculation has 36 channels. That number is 
reduced to 30 channels by eliminating the states with an excitation energy 
larger 10 MeV. A similar Ch-30 calculation is constructed for 
$^{40}$Ca+$^{90}$Zr by eliminating excitations larger than 10.5 MeV.


\subsection{$^{40}$Ca+$^{90}$Zr}

\begin{figure}
\includegraphics[width = 8cm]{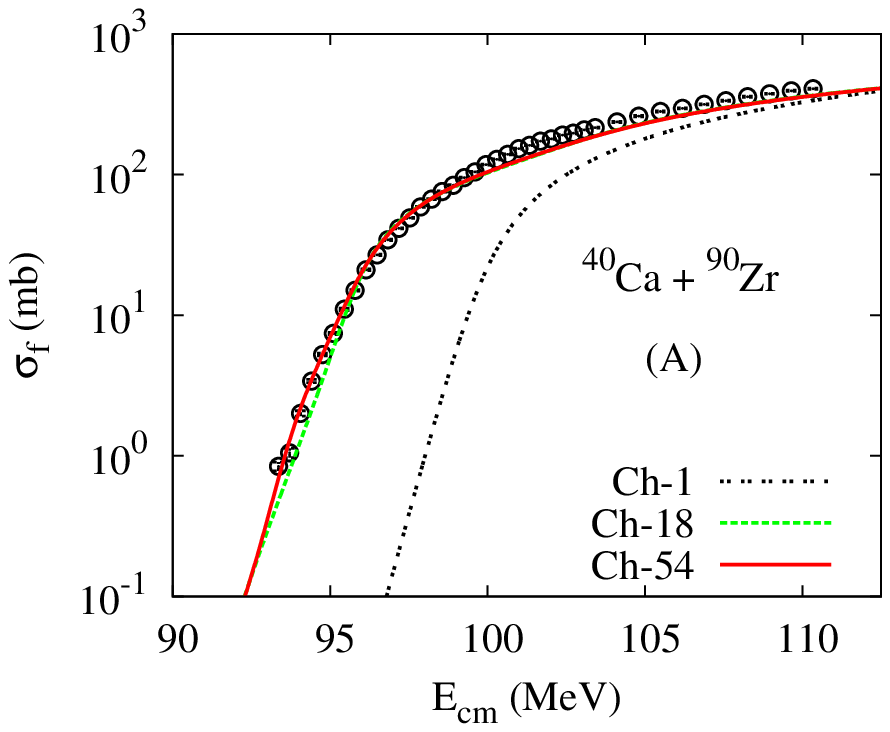}
\includegraphics[width = 8cm]{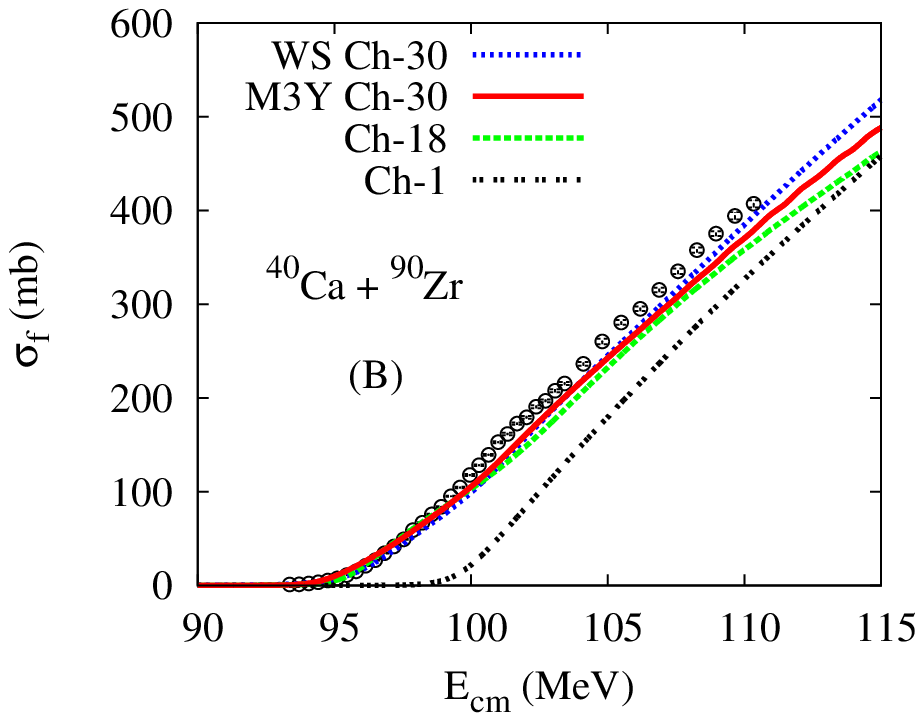}
\caption{(color online)
Logarithmic (A) and linear plot (B) of the measured fusion cross 
sections for $^{40}$Ca+$^{90}$Zr \cite{timca40zr} are compared to 
the Ch-1, Ch-18, Ch-30 and Ch-54 calculations that are based on 
the predicted M3Y+rep potential.  The top curve in the right panel 
is a Ch-30 calculation that is based on a Woods-Saxon potential.}
\label{4090ff}
\end{figure}

The cross sections one obtains for this system in the Ch-18 
coupled-channels calculations are shown by the (green) 
dashed curve in Fig. \ref{4090ff}.
The calculation under-predicts the low-energy data as illustrated 
in Fig. \ref{4090ff}A, and it is also too small at high energies as
shown in Fig. \ref{4090ff}B. One way to reduce the discrepancies with
the data is to include couplings to one- and two-proton transfer channels.  
The effective $Q$ value for one-proton transfer, from the $d_{3/2}$ 
orbit in $^{40}$Ca to the empty $g_{9/2}$ orbit in $^{90}$Zr,  is 
-0.73 MeV (see Table III.). The effective $Q$ value for two-proton 
transfer is +3.05 MeV but it is set equal to +1 MeV in the calculations,
as it was done in the calculations for $^{48}$Ca+$^{90}$Zr, because that 
value provides the best description of the fusion data.  The strength of 
the pair-transfer $\sigma_{2p}$ is adjusted to optimize the fit to the data.

\noindent {\it Influence of transfer.}
The calculation with one- and two-proton transfer that is built on the 
Ch-18 calculation has 54 channels and is shown by the solid curve in 
Fig. \ref{4090ff}A. The optimum strength of the 
pair-transfer is $\sigma_{2p}$ = 0.035 fm, which produces a modest pair 
transfer cross section of 34 mb at 110 MeV. 
Although the transfer improves the fit to the data, in particular at low 
energies, there are still some discrepancies at high energies, where the 
data exceed the calculated cross sections by 10-20\%. 

\noindent {\it Effects of multiphonon excitations.}
The result of the three-phonon calculation Ch-30 is shown by the
solid curve in Fig. \ref{4090ff}B. The calculation is in slightly better 
agreement with the data at high energy but the overall $\chi^2/N$ is 
the same as obtained in the Ch-18 calculation (see Table VI.)
One could also include transfer channels in combination with the Ch-30
multiphonon excitation channels discussed above. 
However, the resulting Ch-90 calculation does not improve the $\chi^2/N$ 
by much (see Table VI.) The reason is that the couplings to transfer 
channels enhance the fusion cross section at low energy but it does not 
have much effect at high energy where the discrepancy with the data is 
the largest.

It is surprising that multiphonon excitations apparently do 
not improve the fit to the $^{40}$Ca+$^{90}$Zr fusion data. 
This statement is based on the observation that the $\chi^2/N$ is 
the same in the Ch-18 and Ch-30 calculations. 
A possible explanation for this result is that the effect of 
multiphonon excitations are exaggerated in the Ch-30 calculation. 
Since the multiphonon excitations of $^{90}$Zr were tested in 
Sect. III.A and found to be reasonable, one could instead question 
the multiphonon excitations of $^{40}$Ca.
The octupole excitation of $^{40}$Ca, for example, is very strong 
but it is possible that the two-phonon excitation of this state is 
not as collective as described by the harmonic oscillator model. 
In fact, it turns out that one can achieve a much better fit to the data
by excluding the two-phonon excitation of the $3^-$ state in $^{40}$Ca.
The basic two-phonon calculation will then have 17 channels (Ch-17) and 
the basic three-phonon calculations will now contain 27 channels (Ch-27). 
The calculation Ch-81 that is built on the Ch-27 calculation and includes
in addition couplings to the one- and two-proton transfer channels, 
provides the best fit to the data with a $\chi^2/N$ = 1.85. 
The best fit is achieved with a modest pair-transfer strength of 
$\sigma_{2p}$ = 0.035 fm.

\begin{figure}
\includegraphics[width = 8cm]{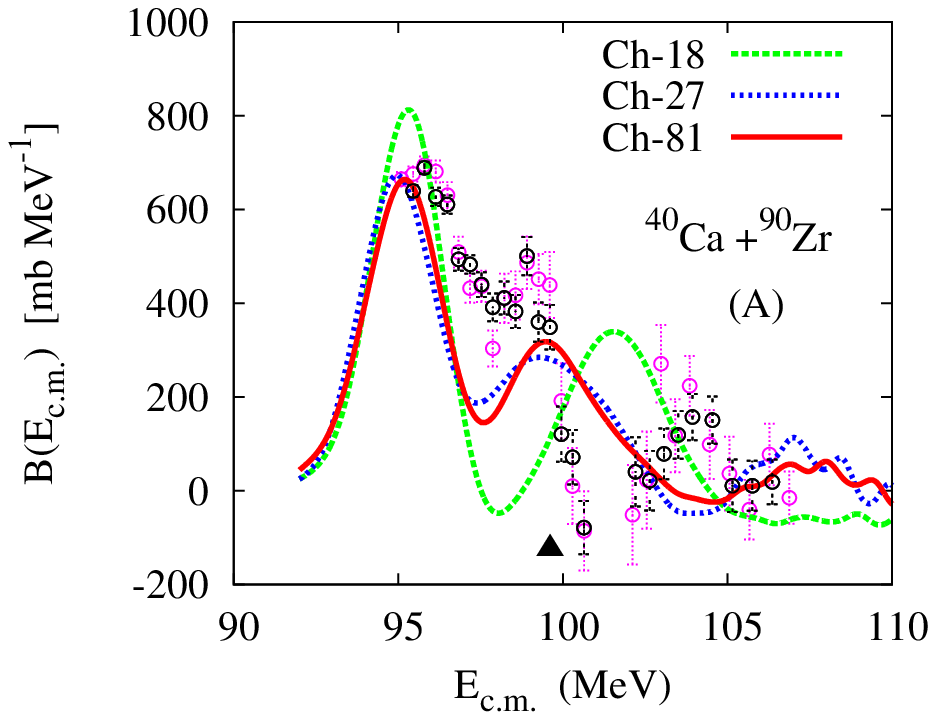}
\includegraphics[width = 8cm]{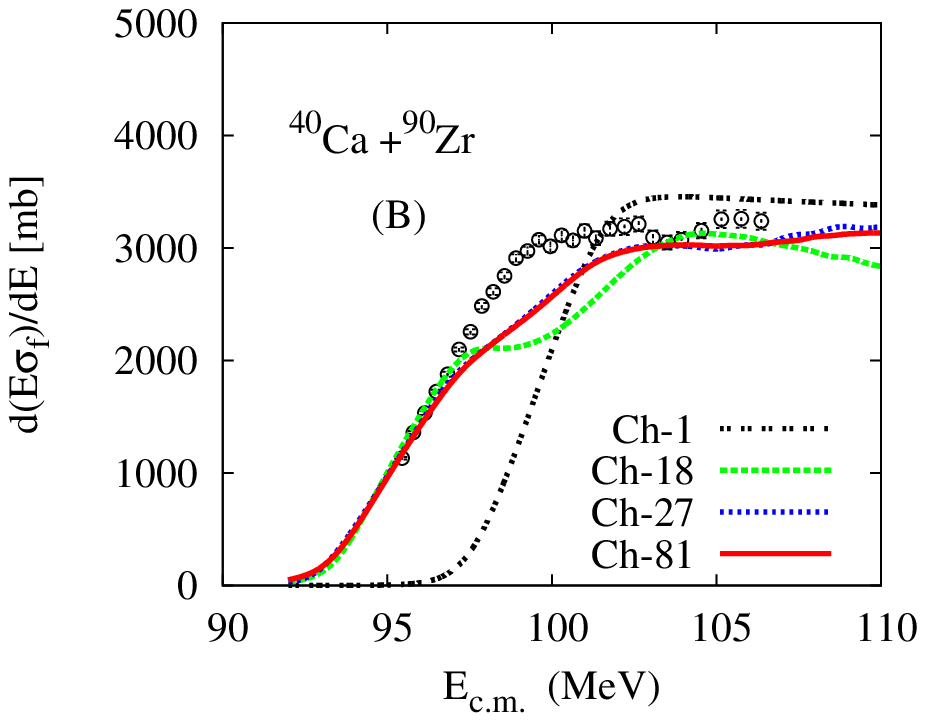}
\caption{Barrier distributions (A) and first derivative (B) of the 
energy-weighted cross sections for the fusion of $^{40}$Ca+$^{90}$Zr.
The coupled-channels calculations use the predicted M3Y+rep potential.
The large triangle in (A) indicates the energy of the Coulomb barrier.}
\label{4090f2d}
\end{figure}

\noindent {\it Derivatives of the cross section.}
The influence of couplings to multiphonon excitations and transfer
channels on the barrier distribution and the first derivative of the 
energy-weighted cross sections is illustrated in Fig. 
\ref{4090f2d}A and \ref{4090f2d}B, respectively.
It is seen that the influence of multiphonon excitations in Ch-27 
calculations improves the agreement with the shape of the measured 
barrier distribution in comparison to the Ch-18 two-phonon calculation 
but some discrepancies remain.  
It is also seen that the influence of transfer, which improves the 
overall $\chi^2/N$ considerably in Ch-81 calculations, has only a 
minor effect on the barrier distribution and on the first 
derivative of the energy-weighted cross section.

The above discussion shows that the predicted ion-ion potential for
$^{40}$Ca+$^{90}$Zr is reasonable and provides a good starting point 
for the analysis of the fusion data. Moreover, the agreement of the 
Ch-81 calculation with the fusion data is satisfactory in view of 
the uncertainties that exist in the nuclear structure input to 
multiphonon excitations.  

It should be pointed out that the fusion data for $^{40}$Ca+$^{90}$Zr 
do not follow the trend that is observed in the fusion of other 
heavy-ion systems.  For example, the high-energy data are enhanced 
compared to calculations that use a standard Woods-Saxon potential.
This is illustrated in Fig. \ref{4090ff}B where the top curve is a 
Ch-30 calculation that used a standard Woods-Saxon potential
with an adjusted radius. 
It is seen that the data exceed this calculation at  high energies. 
This is opposite to the systematics pointed out by Newton et al. 
\cite{newton} who showed that most data sets are suppressed at high 
energies compared to calculations that use a standard Woods-Saxon 
potential of the form proposed in Ref. \cite{AW95}. 

Other calculations that use a standard Woods-Saxon potential 
were also performed. The radius of the potential was adjusted in each 
case to optimize the fit to the data. It turns out that the couplings 
to transfer channels do not improve the fit to the data.
The parameters of the best Ch-18, Ch-30 and Ch-27 calculations are 
shown in Table VI, and it is seen that the smallest $\chi^2/N$ is
achieved in the Ch-27 calculation. The fit is essentially as good
as obtained in the Ch-81 calculation that used the M3Y+rep
potential. The reason is that the smallest measured cross section 
is about 1 mb so the expected fusion hindrance at very low energies
has not yet set in.

\subsection{$^{40}$Ca+$^{96}$Zr}

The fusion data for this system \cite{timca40zr,stef4096} were 
recently analyzed by coupled-channels calculations that used 
a standard Woods Saxon potential. 
The Ch-23 calculations that were performed included one-, two- 
and some three-phonon excitations, and the Ch-69 calculations 
included in addition the couplings to one- and two-neutron 
transfer channels.  
The one-neutron transfer, from the fully occupied $d_{5/2}$ orbit in 
$^{96}$Zr to the unoccupied $f_{7/2}$ state in $^{40}$Ca, has an
effective $Q$ value of +0.61 MeV (see Table III), and the effective 
$Q$ value for two-neutron transfer was set to +1 MeV with a large 
pair-transfer strength of $\sigma_{2p}$ = 0.5 fm.  The radius of the 
Woods-Saxon well ($R$ = 9.60 fm) was chosen to optimize the fit to 
the data above 100 MeV (see Ref. \cite{stef4096} for details.)

\begin{figure}
\includegraphics[width = 8cm]{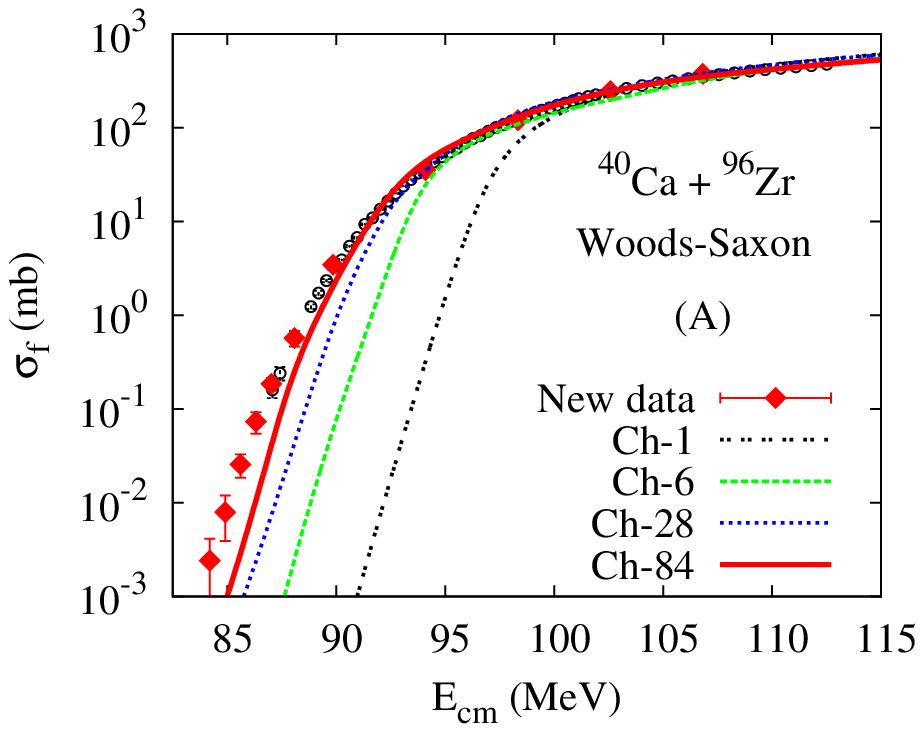}
\includegraphics[width = 8cm]{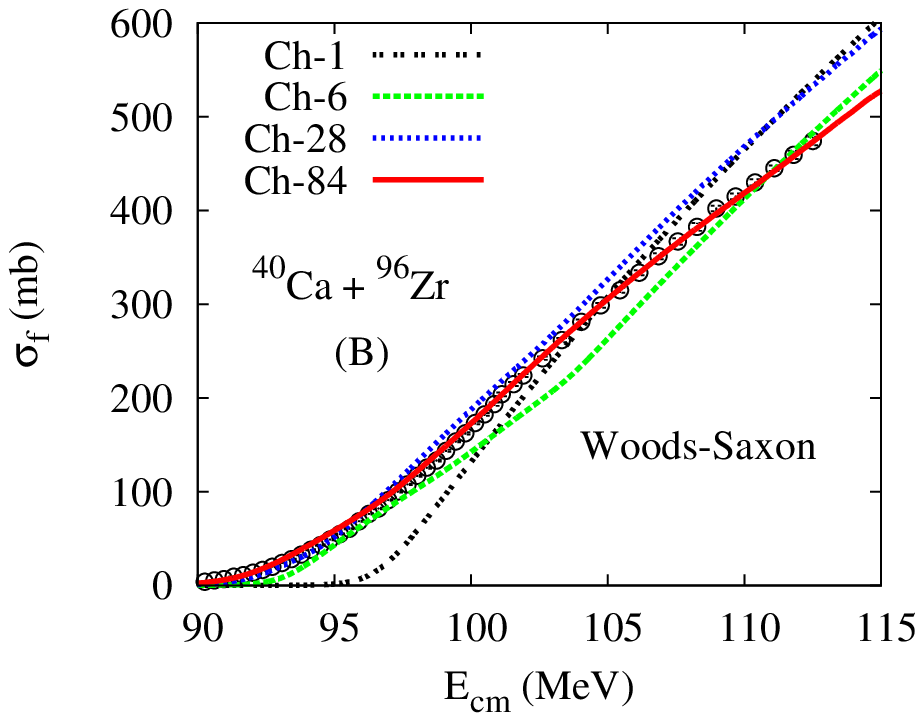}
\caption{(color online)
Logarithmic (A) and linear plot (B) of the measured fusion cross 
sections for $^{40}$Ca+$^{96}$Zr 
(open circles \cite{timca40zr}, solid diamonds \cite{stef4096}). 
They are compared to Ch-1, Ch-6, Ch-28, and Ch-84 calculations that 
are based on a standard Woods-Saxon potential (see Table VII).}
\label{4096ffws}
\end{figure}

Calculations similar to those performed in Ref. \cite{stef4096} 
are repeated here with Ch-28 and Ch-84 calculations that use the same
Woods-Saxon potential. The choice of channels was made because it is
consistent with the best description of the $^{40}$Ca+$^{90}$Zr fusion 
data that was achieved in the previous section with the Ch-27 and Ch-81 
calculations.
The only difference is that the Ch-27 calculation is supplemented with 
the three-phonon excitation of the soft octupole mode in $^{96}$Zr. 
The results of the calculations are presented in Fig. \ref{4096ffws}, 
together with the data of Timmers et al. \cite{timca40zr} and the new 
data by Stefanini et al. \cite{stef4096} that extend the previous 
measurement down to 2.4 $\mu$b. It is seen that the Ch-84 calculation 
shown in Fig.  \ref{4096ffws}B provides an excellent fit to the data 
at high energies. The result is similar to the Ch-69 calculation 
presented in Ref. \cite{stef4096}, and the $\chi^2/N$ of the two 
calculations are essentially the same (see Table VII).

It is interesting to study the sensitivity to the different 
couplings in the high-energy behavior of the calculated cross 
sections. The coupling to one-phonon excitations (Ch-6) reduces 
the cross section compared to the no-coupling calculation (Ch-1). 
The effect of multiphonon excitations in the Ch-28 calculation 
is to enhance the calculated cross section so it exceeds the 
data at high energies. Finally, the coupling to transfer channels 
is so strong in the Ch-84 calculation that it reduces the 
calculated cross section at high energy and brings it into 
agreement with the data. In fact, this behavior was utilized 
in Ref.  \cite{stef4096} to calibrate the strength of the 
pair-transfer coupling.

The Ch-84 calculation shown in Fig. \ref{4096ffws}A underpredicts 
the data the lowest energies. Although it is possible to improve 
the fit to the low-energy data by increasing the strength of the 
pair-transfer coupling, such an increase would cause a reduction 
of the calculated cross section at high energies, which would 
deteriorate to overall agreement with the data.



\begin{figure}
\includegraphics[width = 10cm]{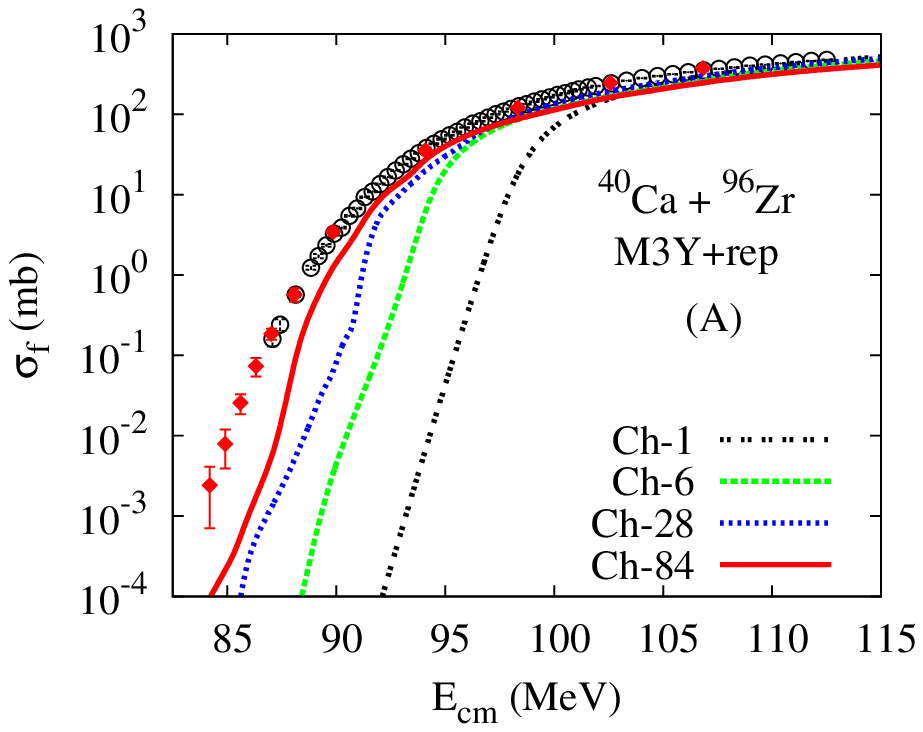}
\includegraphics[width = 10cm]{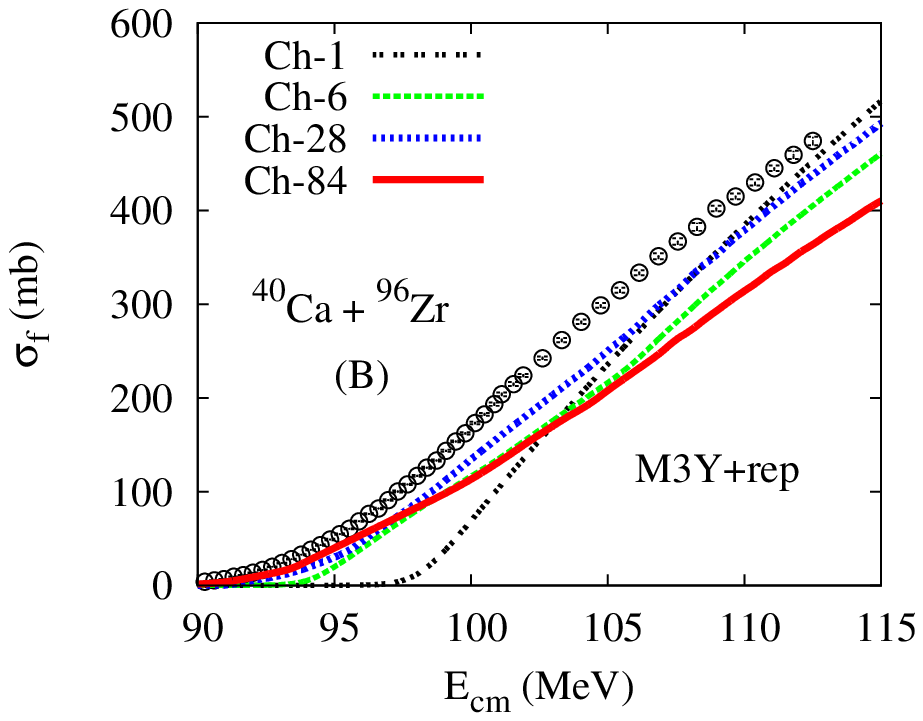}
\caption{(color online)
Logarithmic (A) and linear plot (B) of the measured fusion cross 
sections for $^{40}$Ca+$^{96}$Zr (open circles \cite{timca40zr}, 
solid diamonds \cite{stef4096}). They are compared to Ch-1, Ch-6, Ch-28,
and Ch-84 calculations that use the predicted M3Y+rep potential.}
\label{4096ff}
\end{figure}

The most interesting question is now how well does a Ch-84 calculation  
that is based on the predicted M3Y+repulsion potential agree with 
the data.  The result is shown in Fig. \ref{4096ff}. 
It is seen that the prediction of the data is poor, both at 
low and at high energies. 
The poor result at low energies (see Fig. \ref{4096ff}A) is primarily 
caused by the shallow pocket of the entrance channel potential which 
has a minimum of 87.5 MeV, whereas the data extend down to 84.2 MeV.
The Ch-84 calculation has a threshold near 88 MeV but it 
does extend  to energies below the minimum of the pocket. 
One reason is that fusion can still occur below the minimum of the 
pocket through the pair-transfer channel, which has a positive $Q$ value.  

The failure at high energies of the calculation that is based on 
the predicted M3Y+rep potential is illustrated in the linear plot 
of Fig. \ref{4096ff}B.
The Ch-6 calculation is far below the data 
but the effect of multiphonon excitations in the Ch-28 calculation 
is to enhance the calculated cross section. 
However, the Ch-28 calculation shown in Fig. \ref{4096ff}B is 10-20 \% 
below the data, and this discrepancy increases to 30-50 \% in the 
Ch-84 calculation where the effect of transfer is included.
It is therefore not possible to improve the agreement with data by 
increasing the strength of the pair-transfer. A larger transfer 
strength may improve the calculation at energies below the Coulomb 
barrier but it will reduce the calculated cross section 
even further below the data at high energies.

\begin{figure}
\includegraphics[width = 10cm]{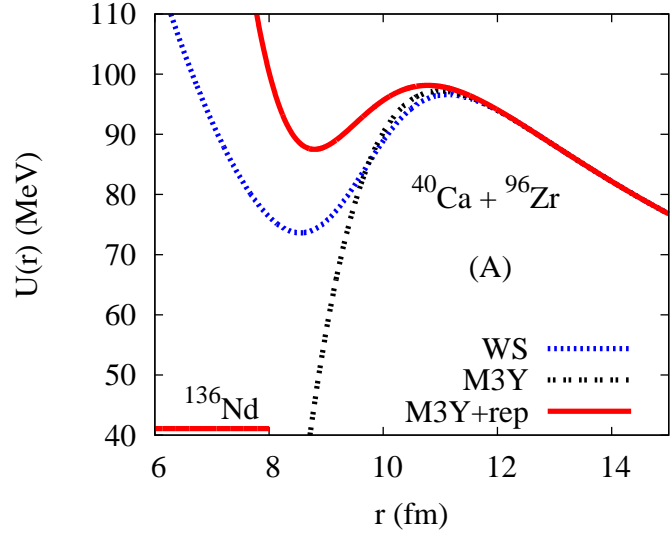}
\caption{(color online)
Entrance channel potentials for $^{40}$Ca+$^{96}$Zr that are based on 
the Woods-Saxon, the M3Y+rep and the pure M3Y nuclear potententials.
The energy of the compound nucleus $^{136}$Nd is indicated.}
\label{4096pot}
\end{figure}

The results of the data analysis are shown in Table VII. 
It is seen that the calculations that use the predicted M3Y+rep 
potential have a $\chi^2/N$ that is much larger than obtained with a
standard Woods-Saxon potential. The main reasons are that the pocket in 
the entrance channel potential for the M3Y+rep interaction is too shallow
($V_{min}$=87.5 MeV) and the Coulomb barrier is too high 
($V_{CBV}$=98.13 MeV) compared to the entrance channel potential of the 
adjusted Woods-Saxon potential. 
These features are illustrated in Fig. \ref{4096pot}, where the entrance 
channel potentials of the different nuclear interactions are compared. 
Another observation in this figure is that the entrance channel 
potential for the pure M3Y interaction is very deep, even deeper than 
the energy of the compound nucleus $^{136}$Nd, but the barrier height 
is slightly larger than obtained with the Woods-Saxon potential.

\begin{figure} 
\includegraphics[width = 10cm]{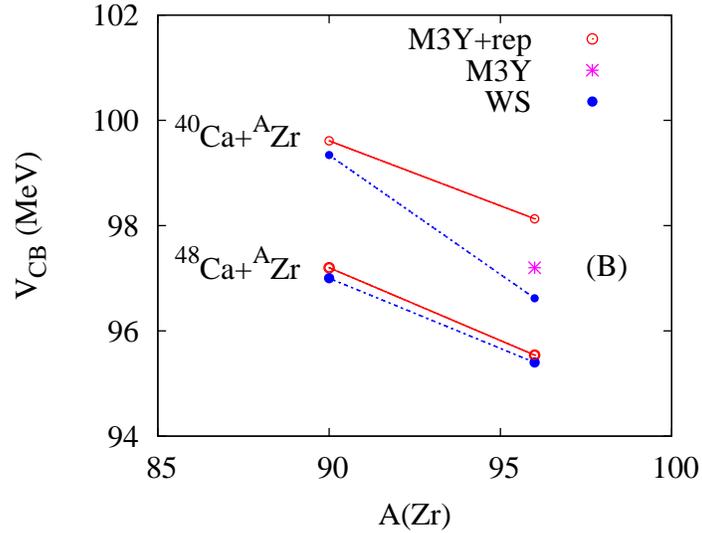}
\caption{(color online)
Heights of the Coulomb barriers for the different Ca+Zr systems.  
Results are shown for the M3Y+rep and the Woods-Saxon potentials. 
The barrier height for the pure M3Y potential 
is also shown for $^{40}$Ca+$^{96}$Zr.}
\label{cazrvcb}
\end{figure}

The systematics of the height of the Coulomb barrier for the four Ca+Zr 
systems and for the two types of potentials considered in this work is
illustrated in Fig. \ref{cazrvcb}. It is seen that the barrier heights
for the Woods-Saxon and the M3Y+rep potentials are almost identical 
for three of the systems but they are different for $^{40}$Ca+$^{96}$Zr.
The M3Y+rep interaction produces a Coulomb barrier that is about 1.5 MeV 
higher than obtained with the Woods-Saxon potential.
The height obtained with the pure M3Y interaction, on the other hand,
is only 0.6 MeV larger than the value obtained with the Woods-Saxon potential. 
This result indicates that one would achieve a better agreement with 
the data by simply ignoring the repulsive part of the double-folding 
potential.  This expectation is confirmed by detailed calculations.
The last two lines of Table VII show the results one obtains for
the pure M3Y interaction, i.~e., for $V_{\rm rep}$=0. It is seen 
that the $\chi^2/N$ is much better than obtained with the M3Y+rep
potential.

The conclusion that the fusion hindrance phenomenon does not occur 
in the fusion of $^{40}$Ca+$^{96}$Zr is supported by the fact that
the adjusted Woods-Saxon, and even the pure M3Y potential provides
a better description  of the data than the M3Y+rep potentia1 does.
The absence of a hindrance at very low energies appears to be 
consistent with the fact that the $Q$ values for pair-transfer are 
large and positive for this system (see Table III). 
The valence nucleons can therefore flow more freely from one nucleus 
to the other without being hindered by Pauli blocking. 
If this interpretation is correct, the disappearance, - or at least 
a reduction, of the repulsive part of the nuclear interaction should 
also occur in reactions of other heavy-ion systems with large positive 
$Q$ values for transfer. This mechanism will affect the isotope 
dependence of the height of the Coulomb barrier and lower it for
systems that have large positive $Q$ values for two-nucleon transfer
reactions.
In this connection it would be desirable to develop a scheme by which
the repulsive part of the nuclear interaction can be calculated  
explicitly by considering the effect of Pauli blocking. 

\section{Conclusion}

We have performed a systematic coupled-channels analysis of the fusion 
data for the four systems, $^{40,48}$Ca+$^{90,96}$Zr, using both a 
standard Woods-Saxon and the M3Y+repulsion, double-folding potential. 
While it is possible to reproduce the data for three of the systems 
in a consistent way by using the M3Y+repulsion potential and a 
realistic nuclear structure input, it is not possible to explain 
the $^{40}$Ca+$^{96}$Zr fusion data within the same framework. 
The data for the latter system are better described by the pure M3Y 
potential or by a standard Woods-Saxon potential.

One of the goals of this work was to test how well the M3Y+rep, 
double-folding potential can be used to predict the ion-ion potential 
once the densities of the reacting nuclei are known.
The nuclear densities of the two calcium nuclei were determined in 
previous analyses of Ca+Ca fusion data. The neutron densities of the 
two zirconium isotopes were determined by analyzing the fusion data 
induced by $^{48}$Ca, whereas the proton densities were constrained 
by electron scattering.  
The neutron skin thickness of each of the two zirconium isotopes
extracted from the analysis of the fusion data is consistent with 
the results of anti-proton experiments.  This is a nice consistency
check of the coupled-channels calculations.

The predicting power of the double-folding method was tested by 
calculating the M3Y+repulsion potential for the $^{40}$Ca induced 
reactions with the two zirconium isotopes and analyzing the fusion 
data with coupled-channels calculations. 
This approach worked fairly well for $^{40}$Ca+$^{90}$Zr 
but it failed for $^{40}$Ca+$^{96}$Zr, primarily 
because the predicted Coulomb barrier is too high 
and the pocket in the entrance channel potential is too shallow. 

Although the influence of transfer plays a role in explaining the 
fusion data for most of the Ca+Zr systems, it is only in the case 
of $^{40}$Ca+$^{96}$Zr that transfer is expected to have a major
impact. This feature was recognized in the original work of Timmers 
et al. \cite{timca40zr}, where the effect of transfer was expected 
to be the reason for the large enhancement of the measured subbarrier 
fusion cross sections.  
The surprising new result 
is that the ion-ion potential predicted for this system by the M3Y+rep
interaction is unrealistic. 
The failure of the prediction is ascribed to the influence of transfer 
reactions which is expected to be particularly strong because the 
transfer $Q$ values are large and positive and the transfer can 
therefore occur without the hindrance imposed by Pauli blocking. 

The repulsive part of the M3Y+repulsion potential, which explains
the hindrance phenomenon observed in the fusion of many heavy-ion 
systems at extreme subbarrier energies, is usually calibrated to 
produce a realistic nuclear incompressibility for 
overlapping nuclei. 
The present work suggests that the fusion hindrance and the 
hindrance of transfer reactions imposed by the Pauli blocking 
are somehow related, because they are both absent in reactions 
of $^{40}$Ca+$^{96}$Zr. In this connection it would be very 
attractive if one could calculate the repulsive part of the 
ion-ion potential by considering the effect of Pauli blocking
explicitly.

The analysis of the different data sets revealed a number of 
interesting problems.
For example, the fusion data for 
$^{48}$Ca+$^{90}$Zr are suppressed at high energies compared to 
calculations that use a standard Woods-Saxon potential, but this 
discrepancy was removed by applying the M3Y+rep potential. 
Another problem is that some of the data sets are enhanced at high 
energies compared to coupled-channels calculations that include  
couplings to one- and two-phonon excitations. The problem was 
resolved for $^{48}$Ca+$^{96}$Zr 
by considering the influence of three-phonon excitations.
Finally, the calculated barrier distribution for fusion consists
typically of a few strong peaks, whereas the measured distribution 
is sometimes broad and smooth. This difference may be caused by 
the simplified models of excitations and transfer that
are used in the calculations.

A good explanation of the $^{40}$Ca+$^{96}$Zr fusion data is still 
missing. We have shown that it is not possible to predict the ion-ion 
potential reliably for this system by the double-folding technique.
It appears that an adjusted Woods-Saxon potential, or even the pure 
M3Y potential, provides a much more realistic description.  However, 
none of the calculations can account for the data at the lowest energies.
A clear improvement of the coupled-channels calculations would 
be to calibrate the transfer couplings, in particular for the 
two-nucleon transfer, so that the transfer data were reproduced 
by the calculations.  This approach is currently being pursued. 
Another possibility is to apply a more realistic ion-ion potentials 
in the exit channels.  This idea was proposed by Sargsyan et al. 
\cite{sargsyan}, who used ion-ion potentials in the exit channels 
for transfer reactions that are different from the entrance channel 
potential because of deformation effects.  
Both approaches are worth pursuing.

{\bf Acknowledgments}.
We acknowledge fruitful discussions with B. B. Back and  G. Montagnoli.
This work was supported by the U.S. Department of Energy, 
Office of Nuclear Physics, Contract No. DE-AC02-06CH11357.
The research leading to these results has received funding from
the European Union Seventh Framework Programme FP7/2007-2013 under Grant
Agreement No. 262010-ENSAR.

\begin{table}
\caption{Adopted structure of the excited states in 
$^{40}$Ca \cite{flem,mon4040}, $^{48}$Ca \cite{flem,esb4848},
and $^{90,96}$Zr \cite{NDS,spear02}.
Note the $3^-$ and $5^-$ excitations in $^{90}$Zr are combined 
into one effective $3^-$ state.}
\begin{tabular} {|c|c|c|c|c|c|c|}
\colrule
Nucleus & $\lambda^\pi$ &  E$_x$ (MeV) & B(E$\lambda$) (W.u.) &  
$\beta_\lambda^C$ & $\sigma_\lambda^C$ (fm) 
& $\sigma_\lambda^N$ (fm) \\
\colrule
 $^{40}$Ca  & $2^+$     & 3.904 & 2.26(14)& 0.119 & 0.138 & 0.125 \\
            & $3^-$     & 3.737 & 27(4)   & 0.402 & 0.465 & 0.315 \\
            & $5^-$     & 4.491 & 16      & 0.297 & 0.344 & 0.175 \\
\colrule
$^{48}$Ca   & $2^+$       & 3.832 & 1.71(9)& 0.102 & 0.126 & 0.190 \\
            & $3^-$       & 4.507 & 5.0(8) & 0.203 & 0.250 & 0.190 \\
 ignore:    & $5^-$       & 5.146 & 0.3    & 0.040 & 0.049 & 0.038 \\
\colrule
 $^{90}$Zr  & $2^+$      & 2.186 &5.37(20)  & 0.092 & 0.140 & 0.140   \\
            & $5^-$      & 2.319 & 8.7(4)   & 0.108 & 0.164 & 0.164   \\
Ref. \cite{spear02} & $3^-$      & 2.748 & 28.9(15) & 0.210 & 0.319 & 0.319   \\
effective:  & $3^-$ & 2.658 &          & 0.236 & 0.358 & 0.358   \\
\colrule
 $^{96}$Zr  & $2^+$      & 1.751 & 4(3)  & 0.079 & 0.123 & 0.123 \\
            & $3^-$      & 1.897 & 57(4) & 0.295 & 0.457 & 0.457 \\
\colrule
\end{tabular}
\end{table}

\begin{table}
\caption{Parameters of the densities in $^{40,48}$Ca and $^{90,96}$Zr:
the diffuseness $a_r$ associated with the repulsion (Eq. (7)),
the radius $R$, the diffuseness $a$, and the last column is the
rms radius.  The matter density parameters for the calcium isotopes 
were determined previously \cite{mon4040,esb4848}. The second column 
shows the fusion reaction that was used to determine the parameters.  
The parameters for the proton (p) densities in $^{90}$Zr and $^{96}$Zr 
reproduce the point-proton rms radii, $\langle r^2\rangle_{\rm pp}$ = 
4.198 and 4.281 fm, respectively, extracted from the measured charge 
radii \cite{angeli}.  The parameters for the neutron (n) densities 
in $^{90}$Zr and $^{96}$Zr are from Tables IV and V.}
\begin{tabular} {|c|c|c|c|c|c|}
\colrule
Nucleus & Reaction & $a_r$ (fm) & 
 $R$ (fm) & $a$ (fm) & $\langle r^2\rangle^{1/2}$ (fm) \\
\colrule
$^{40}$Ca & $^{40}$Ca+$^{40}$Ca \cite{mon4040} & 0.42 & 3.47  & 0.56 & 3.400  \\ 
$^{48}$Ca & $^{48}$Ca+$^{48}$Ca \cite{esb4848} & 0.43 & 3.798 & 0.54 & 3.562  \\ 
\colrule
$^{90}$Zr (p) &   &  & 4.72  & 0.56 & 4.207 \\ 
$^{90}$Zr (n) & $^{48}$Ca+$^{90}$Zr Ch-15 & 0.40 & 4.925 & 0.56 & 4.346 \\ 
$^{90}$Zr (n) & $^{48}$Ca+$^{90}$Zr Ch-45 & 0.39 & 4.835 & 0.56 & 4.285 \\ 
\colrule
$^{96}$Zr (p) &  &  & 4.86 & 0.55 & 4.284  \\ 
$^{96}$Zr (n) & $^{48}$Ca+$^{96}$Zr Ch-15 & 0.395 & 5.20 & 0.55 & 4.517  \\  
$^{96}$Zr (n) & $^{48}$Ca+$^{96}$Zr Ch-72 & 0.395 & 5.10 & 0.55 & 4.448  \\  
\colrule
\end{tabular}
\end{table}

\begin{table}
\caption{Effective $Q$ values (in MeV) for the most favorable one-nucleon 
and two-nucleon transfer reactions in $^{40}$Ca+$^{90,96}$Zr and 
$^{48}$Ca+$^{90,96}$Zr collisions, and the adopted strength $\sigma_{2p}$ 
(in fm) of the pair transfer.} 
\begin{tabular} {|c|c|c|c|c|c|}
\colrule
System  & $Q_{1n}$ & $Q_{2n}$ & $Q_{1p}$ & $Q_{2p}$ & $\sigma_{2p}$ \\
\colrule
$^{48}$Ca+$^{96}$Zr &  -2.64 & -2.67 & -4.07 & -3.49 & 0.05 \\
$^{48}$Ca+$^{90}$Zr &  -2.83 & -1.55 & -0.92 & +2.22 & 0.15 \\
$^{40}$Ca+$^{90}$Zr &  -3.50 & -1.25 & -0.73 & +3.05 & 0.035 \\
$^{40}$Ca+$^{96}$Zr &  +0.61 & +5.73 & +1.55 & +7.63 & 0.50 \\
\colrule
\end{tabular}
\end{table}

\begin{table}
\caption{Best fits to the $^{48}$Ca+$^{90}$Zr fusion data of Ref. 
\cite{stefca48zr}.  The $\chi^2/N$ (last column) includes a 5\% 
systematic error.  The 1st column shows the type of calculation.
The 2nd column indicates the type of ion-ion potential, either the
Woods-Saxon (WS) potential or the diffuseness $a_r$ of the $^{90}$Zr 
density associated with the repulsive part of the M3Y+rep potential. 
The 3rd column is the radius, either of the WS potential or  
the neutron density in $^{90}$Zr. 
The radius of the proton density was set to 4.72 fm.
The height of the Coulomb barrier is listed in the 4th column.}
\begin{tabular} {|c|c|c|c|c|}
\colrule
Calculation  & $a_r$ (fm) & $R$ (fm) & $V_{CB}$ (MeV) 
& $\chi^2/N$ \\
\colrule 
 Ch-15 & WS  & 9.539 &  96.89  &  3.59 \\ 
 Ch-45 & WS  & 9.527 &  97.00  &  1.06 \\ 
\colrule
 Ch-15 & 0.40 & 4.925 & 97.02 &  3.12 \\ 
 Ch-23 & 0.40 & 4.925 & 97.02 &  2.30 \\ 
 Ch-45 & 0.39 & 4.835 & 97.20 &  0.69 \\ 
 Ch-69 & 0.39 & 4.835 &  97.20 &  0.69 \\ 
\colrule 
\end{tabular}
\end{table}

\begin{table}
\caption{Best fits to the $^{48}$Ca+$^{96}$Zr fusion data of Ref. \cite{stefca48zr}.
The $\chi^2/N$ (last column) includes a 5\% systematic error.
The 1st column shows the type of calculation.
The 2nd column indicates the type of ion-ion potential, either the
Woods-Saxon (WS) or the diffuseness $a_r$ of the $^{96}$Zr density 
associated with the repulsive part of the M3Y+rep potential.  
The 3rd column is the radius, either of the WS potential  or the 
neutron density in $^{96}$Zr. The radius of the proton
density was set to 4.86 fm. 
The height of the Coulomb barrier is listed in the 4th column.}
\begin{tabular} {|c|c|c|c|c|}
\colrule
 Calculation  & $a_r$ (fm) & $R$ (fm) & $V_{CB}$ (MeV) 
& $\chi^2/N$ \\
\colrule
 Ch-15   & WS & 9.689 &  95.40 & 1.53 \\  
 Ch-24   & WS & 9.671 &  95.56 & 3.43 \\ 
\colrule
 Ch-15 &  0.395 & 5.20  & 95.54 & 1.49 \\
 Ch-23 &  0.395 & 5.20  & 95.54 & 2.16 \\
 Ch-24 &  0.395 & 5.10  & 95.86 & 2.22 \\
 Ch-72 &  0.395 & 5.10  &  95.86 & 0.76 \\
\colrule
\end{tabular}
\end{table}

\begin{table}
\caption{Analysis of the $^{40}$Ca+$^{90}$Zr fusion data \cite{stefca48zr}.
The $\chi^2/N$ (last column) includes a 7\% systematic error.
The type of calculation is listed in the 1st column. 
The 2nd column is the adjusted radius of the WS potential, 
or the radius of the neutron density in $^{90}$Zr 
obtained in the Ch-69, M3Y+rep calculation of Table IV.
The 3rd and 4th columns show the minimum of the pocket and the height 
of the Coulomb barrier in the entrance channel potential.}
\begin{tabular} {|c|c|c|c|c|}
\colrule
Calculation & $R$ (fm) & $V_{min}$ (MeV) & $V_{CB}$ (MeV) &
 $\chi^2/N$ \\ 
\colrule 
WS Ch-18 & 9.312 & 78.91 & 99.49 & 2.60 \\ 
WS Ch-30 & 9.302 & 79.08 & 99.58 & 3.92 \\ 
Ws Ch-27 & 9.327 & 78.66 & 99.34 & 1.65 \\ 
\colrule 
 Ch-18 & 4.835 & 86.15 & 99.61 &  4.94 \\ 
 Ch-30 & 4.835 & 86.15 & 99.61 &  4.94 \\ 
 Ch-27 & 4.835 & 86.15 & 99.61 &  3.37 \\ 
\colrule
 Ch-54 & 4.835 & 86.15 & 99.61 &  2.53 \\ 
 Ch-90 & 4.835 & 86.15 & 99.61 &  4.25 \\ 
 Ch-81 & 4.835 & 86.15 & 99.61 &  1.85 \\ 
\colrule
\end{tabular}
\end{table}

\begin{table}
\caption{Analysis of the $^{40}$Ca+$^{96}$Zr fusion data \cite{stefca48zr}.
The $\chi^2/N$ in the (last column) includes a 7 \% systematic error.
The type of calculation is listed in the 1st column.
The 2nd column is the adjusted radius of the WS potential, or the radius
of neutron density in $^{96}$Zr obtained in the Ch-72, M3Y+rep calculation 
of Table V.
The 3rd and 4th columns show the minimum of the pocket and the height 
of the Coulomb barrier in the entrance channel potential.
The last two lines show the results of Ch-69 and Ch-84 calculations 
that use the pure M3Y potential, i.~e., for $V_{\rm rep}$=0.}
\begin{tabular} {|c|c|c|c|c|}
\colrule
Calculation & $R$ (fm) & $V_{min}$ (MeV) & $V_{CB}$ (MeV) &
 $\chi^2/N$  \\ 
\colrule 
WS Ch-69 & 9.599 & 73.62 & 96.62 & 4.0 \\ 
WS Ch-84 & 9.599 & 73.62 & 96.62 & 4.1 \\ 
\colrule 
Ch-69 & 5.10 & 87.5 & 98.13 & 22 \\ 
Ch-84 & 5.10 & 87.5 & 98.13 & 23 \\ 
\colrule 
Ch-69 $V_{\rm rep}$ = 0 & 5.10 & -311 & 97.20 & 5.3 \\ 
Ch-84 $V_{\rm rep}$ = 0 & 5.10 & -311 & 97.20 & 6.0   \\ 
\colrule
\end{tabular}
\end{table}


\end{document}